\documentclass{ar-1col}
\usepackage{arydshln}
\usepackage[numbers]{natbib}
\usepackage{bm}
\usepackage{url}
\usepackage{amsmath,amssymb}
\usepackage{pifont}
\usepackage{subfigure}
\usepackage[dvipsnames]{xcolor}
\usepackage{multirow}

\def\ie{{\it i.e.\/}}
\def\etc{{\it etc.\/}}
\def\eg{{\it e.g.\/}}
\def\frac#1#2{{\textstyle{#1 \over #2}}}
\def\half{\frac{1}{2}}
\def\yd{^\dagger}
\def\nd{^{\vphantom{\dagger}}}
\def\np{^{\vphantom{\prime}}}
\def\uar{\uparrow}
\def\dar{\downarrow}
\def\nhat{{\hat n}}
\def\tt{\textsf{t}}
\def\ij{{\langle ij\rangle}}
\def\RPsi{{\rm\Psi}}
\def\sigb{{\bar\sigma}}
\def\ism{{i\sigma m}
\def\itm{{i\sigb m}}
\def\cbar{{\bar c}}
\def\ve{\varepsilon}
\def\vchi{\vec{\chi}}}
\def\eps{\epsilon}
\def\ssp{{\sigma\sigma'}}
\def\ttp{{\alpha\beta}}
\def\Dbar{{\bar\Delta}}

\def\vF{v\nd_{\rm F}}
\def\kF{k\nd_{\rm F}}
\def\gtwid{\,{\raise.3ex\hbox{$>$\kern-.75em\lower1ex\hbox{$\sim$}}}\,}

\def\CO{{\cal O}}
\def\CR{{\cal R}}
\def\CU{{\cal U}}

\def\Bk{\bm{k}}
\def\Br{\bm{r}}
\def\Bp{\bm{p}}

\def\BS{\bm{S}}
\def\BQ{\bm{Q}}
\def\BR{\bm{R}}
\def\Bsigma{\bm{\sigma}}
\def\Bchi{\bm{\chi}}
\def\cbar{{\bar c}}

\def\HBk{\hat{\bm{k}}}

\def\HBk{\Omega\nd_{\bm{k}}}

\def\BkF{\Bk\np_{\rm F}}
\def\BkpF{\Bk'_{\rm F}}
\def\BpF{\Bp\np_{\rm F}}

\def\Rep{\textsf{Re}}
\def\Imp{\textsf{Im}}
\def\pz{\partial}
\def\Tr{\textsf{Tr}}

\def\ket#1{{\big| \, #1\, \big\rangle}}

\def\sket#1{{|  \, #1 \,  \rangle}}

\def\sexpect#1#2#3{{\langle \, #1 \, | \,  #2  \, | \, #3 \, \rangle}}
\def\tSU{\textsf{SU}}
\def\tO{\textsf{O}}
\def\SA{\textsf{A}}
\def\SB{\textsf{B}}
\def\tU{\textsf{U}}
\def\tSO{\textsf{SO}}
\def\tSp{\textsf{Sp}}

\def\Ht{{\hat t}}
\def\HH{{\hat H}}
\def\HN{{\hat N}}
\def\HK{{\hat K}}
\def\HS{{\hat S}}
\def\HJ{{\hat J}}

\def\HBS{{\hat \BS}}

\def\MZ{{\mathbb Z}}

\def\tcr{\textcolor{red}}
\def\tcfg{\textcolor{ForestGreen}}
\newcommand{\cmark}{\ding{51}}%
\newcommand{\xmark}{\ding{55}}
\def\gcmk{\tcfg{\cmark}}
\def\rxmk{\tcr{\xmark}}

\jname{Submitted to Annual Review of Condensed Matter Physics}
\jyear{2020}

\begin{document} 

\markboth{Arovas et al.}{The Hubbard Model}

\title{The Hubbard Model}
\author{Daniel P. Arovas$^1$, Erez Berg$^2$, \\ Steven A. Kivelson$^3$, Srinivas Raghu$^{3,4}$
\affil{$^1$Department of Physics, University of California at San Diego, La Jolla, California 92093, USA}
\affil{$^2$Department of Condensed Matter Physics, Weizmann Institute of Science, Rehovot 76100, Israel}
\affil{$^3$Department of Physics, Stanford University, Stanford, USA, 94305}
\affil{$^4$Stanford Institute for Materials and Energy Sciences, SLAC National Accelerator Laboratory and Stanford University, Menlo Park, CA 94025, USA}}

\begin{abstract}
The repulsive Hubbard model has been immensely useful in understanding strongly correlated electron systems, and serves as the paradigmatic model of the field. Despite its simplicity, it exhibits a strikingly rich phenomenology which is reminiscent of that observed in quantum materials. Nevertheless, much of its phase diagram remains controversial. Here, we review a subset of what is known about the Hubbard model, based on exact results or controlled approximate solutions in various limits, for which there is a suitable small parameter. Our primary focus is on the ground state properties of the system on various lattices in two spatial dimensions, although both lower and higher dimensions are discussed as well. Finally, we highlight some of the important outstanding open questions.
\end{abstract}

\begin{keywords}
Hubbard model, Strongly correlated electrons, quantum materials, unconventional superconductivity, magnetism
\end{keywords}
\maketitle

\tableofcontents

\section{Introduction}

The Hubbard model~\cite{Hubbard1963} has come to play a paradigmatic role in the theory of electronic correlations in quantum materials where interactions play an essential role.  Its centrality in the quantum statistical mechanics of interacting fermions is comparable to that of the Ising model in classical statistical mechanics, yet almost sixty years after it was first described, there remain unsettled basic questions, even concerning the actual phases that arise for various lattice geometries. 
Conversely, the more progress that is made in obtaining theoretical solutions, the clearer it becomes that this simple model can exhibit a startling array of phases and regimes, many of which have clear parallels with observed behaviors of a wide variety of complex materials.  For instance, there is compelling evidence that ferromagnetism, various forms of antiferromagnetism, unconventional superconductivity, charge-density waves, electronic liquid crystalline phases, and  topologically ordered phases (\eg, ``spin liquids''), among other phases, occur in specific realizations of the Hubbard model.

It is our purpose here to summarize, to the extent possible in a brief article, what is {\em established\/} concerning the quantum phases of the Hubbard model.  While some discussion of the model on small clusters and in one dimension ($d=1$) is included for pedagogic purposes, the focus is primarily on the model on regular lattices (\ie\ in the absence of disorder) in $d=2$ and $d=3$. Likewise while we include some discussion of finite temperature results, our focus is on ground state properties, mostly to keep the scope of the article manageable.  In the strong coupling limit and in the special case of one electron per site, the Hubbard model reduces to an insulating quantum Heisenberg antiferromagnet, which itself can exhibit many different and fascinating quantum phases; these will be covered to an extent but mostly we will focus on phases that occur for generic electron densities. Furthermore, we have mostly focused on equilibrium  properties of the model, especially those essential thermodynamic correlation functions that characterize distinct quantum phases of matter.

Even within this limited scope, we have restricted ourselves to ``controlled'' solutions - by which we mean either exact results (typically obtained using numerical methods of various sorts) or in cases where there is an approximation scheme that becomes asymptotically exact as an explicitly identified small parameter tends to zero.  This has led us to omit discussion of a variety of new and powerful ``mean-field'' methods, including dynamical mean field theory and dynamical cluster approximations, which have profitably been applied to this problem. Fortunately, a companion paper to the present paper~\cite{gul} has these approaches as its complementary focus.

We also will not explore the relation between the Hubbard model and experiments in any particular quantum material. Nonetheless, much of the modern resurgence of interest in this model is a consequence of the role it has played in the study of high temperature superconductivity in the cuprates.  Originally, this connection was suggested on rather basic phenomenological grounds~\cite{anderson87,emery87} - the ``parent state'' in the cuprates is a quasi-2d antiferromagnetic insulator, and the Hubbard model on the square lattice is the simplest possible model of a doped antiferromagnet.  It is thus striking that many of the properties of this model that have been uncovered in subsequent theoretical studies turn out to resemble essential features of the electronic states of the cuprates that have been revealed by experiments, {\it viz.\/}: 1) The Hubbard model (at least for a range of $U$ not too large) exhibits a ``dome'' of $d$-wave superconductivity throughout a range of doping.  2) It exhibits antiferromagnetic long-range order when undoped.  3) It evolves to an effectively weakly-coupled Fermi liquid state upon heavy doping.  4) It exhibits a tendency toward formation of long-period unidirectional charge and spin density wave orders (stripes) in the same range of doping in which superconducting correlations (at least locally) appear strongest.  Indeed, as in the materials themselves, the relation between density-wave and superconducting orders appears more complicated than the collocation ``competing orders'' might suggest, leading to their description as ``intertwined.''  As has been noted earlier~\cite{scalhubbard}, this congruence in behavior is sufficiently striking that it encourages the view that the solution of the Hubbard model can, in some sense, be considered to be the solution of the high $T\nd_{\rm c}$ problem.

\section{Theorems}
\label{theorems}
The Hubbard model describes itinerant, interacting electrons of spin-$\half$ hopping on a set $\Lambda$ of 
spatially localized orbitals.  The Hamiltonian is written as
\begin{equation}
\HH=-\sum_{i,j\in\Lambda}\sum_\sigma t\nd_{ij}\,c\yd_{i\sigma}\,c\nd_{j\sigma} + 
U\sum_{i\in\Lambda} c\yd_{i\uar} c\yd_{i\dar} c\nd_{i\dar} c\nd_{i\uar}\quad,
\end{equation}
where the hopping $t_{ij}$ is often restricted to nearest-neighbor sites, \ie\ $t\nd_{ij}=t^*_{ji}=t\,\delta\nd_{|i-j|,1}$, but generalizations can include further neighbor hopping and/or Peierls phase factors to account for nonzero magnetic flux.  Unless explicitly noted, we will confine our attention to the zero flux (time-reversal invariant) case. 
Moreover, we will consider only regular, connected lattices, \ie\ we do not consider the effects of ``disorder.''
The electron filling is defined to be $n\equiv N/|\Lambda|$ where $|\Lambda|$ is the total number of sites and $N$ is the number of electrons.  The half-filled band, with one electron per site, corresponds to $n=1$.

{\sl Continuous global symmetries} -- The model's most apparent symmetry is a global $\tU(2)$, under which $c\nd_{i\sigma}\to \CU\nd_{\sigma\sigma'}\,c\nd_{i\sigma'}$ with $\CU\in\tU(2)$ the same matrix on all sites $i$.  Separating $\tU(2)=\tU(1)\times\tSU(2)$, the global $\tU(1)$ invariance reflects global charge conservation, hence the total particle number $\HN=\sum_{i,\sigma} c\yd_{i\sigma} c\nd_{i\sigma}$ is a good quantum number. This symmetry may be spontaneously broken in a superconducting state.  The global $\tSU(2)$ invariance reflects spin isotropy, hence $\HS^z$ and $\HBS^2$ are good quantum numbers, with $\HBS=\half\sum_{i,\mu,\nu} c\yd_{i\mu}\,\Bsigma\nd_{\mu\nu}\,c\nd_{i\nu}$.  Global $\tSU(2)$ invariance of the ground state is spontaneously broken in any ferromagnetic or antiferromagnetic state.

{\sl Particle-hole symmetry} -- On a bipartite lattice, the transformation $c\nd_{i\sigma}\to \eta\nd_i\,c\yd_{i\sigma}$, where $\eta\nd_i=\pm$ on alternate sublattices, results in the transformed
Hamiltonian
\begin{equation}
\HH'=-\sum_{i,j\in\Lambda}\sum_\sigma t\nd_{ij}\,c\yd_{i\sigma}\,c\nd_{j\sigma} + 
U\sum_{i\in\Lambda} c\yd_{i\uar} c\yd_{i\dar} c\nd_{i\dar} c\nd_{i\uar} + U\big(|\Lambda|-\HN\big)\quad,
\end{equation}
At half-filling, $\HN=|\Lambda|$, the model is particle-hole symmetric.  In such cases, the system is proven~\cite{LLM1993} to be of uniform density, with $\langle c\yd_{i\sigma} c\nd_{j\sigma}\rangle=\half\,\delta\nd_{ij}$ if $i$ and $j$ are on the same sublattice (with no implicit sum on $\sigma$)\footnote{The conditions for uniform density from ref.~\cite{LLM1993} are much more general, and include the possibility of site-dependent and even nonlocal interactions.}.  However, this does not preclude either magnetic or charge density wave order in the thermodynamic limit.

The \tSU(2) spin symmetry of the Hubbard model is generated by the operators
\begin{equation}
\HS^+=\sum_i c\yd_{i\uar}\, c\nd_{i\dar} \quad,\quad 
\HS^-=(\HS^+)\yd \quad,\quad \HS^z={1\over 2}\,\sum_i\big( c\yd_{i\uar} c\nd_{i\uar} - c\yd_{i\dar} c\nd_{i\dar}\big)\quad,
\end{equation}
all of which commute with $\HH$ as well as with the number operator $\HN=\sum_{i,\sigma} c\yd_{i\sigma} c\nd_{i\sigma}$.  A second, ``hidden'' $\tSU(2)$ symmetry~\citep{Yang1989,Zhang1991} is present on bipartite lattices, and is
generated by the pseudospin operators
\begin{equation}
\HJ^+=\sum_i \eta\nd_i\, c\yd_{i\uar} c\yd_{i\dar} \quad,\quad \HJ^-=(\HJ^+)\yd \quad,\quad
\HJ^z={1\over 2}\,\big(\HN-|\Lambda|\big)\quad,
\end{equation}
where $\eta\nd_i=\pm 1$ on the \SA\ and \SB\ sublattices, respectively.
These operators also satisfy the \tSU(2) algebra $\big[J^\alpha,J^\beta\big]=i\eps\nd_{\alpha\beta\gamma} J^\gamma$.  While the $J^\alpha$'s do not commute with $\HH$ and $\HN$ as do the generators $S^\alpha$ for
physical spin, they are eigenoperators in that
\begin{equation}
\big[\HK,\HJ^\pm\big]=\pm(U-2\mu)\,J^\pm
\end{equation}
and $[\HK,\HJ^z]=0$, where $\HK=\HH-\mu\HN$ is the grand canonical Hamiltonian ($\mu$ is the chemical potential).  Thus, at half-filling, when $\mu=\half U$, the Hubbard model has an additional \tSU(2) symmetry. Hence, the global symmetry group at half-filling is $\tSO(4)\cong\tSU(2)\times\tSU(2)/\MZ\nd_2$, where the $\MZ\nd_2$ is associated with the fact that $\hat{J}^z + \hat{S}^z$ has to be an integer when $|\Lambda|$ is even, and half-integer when $|\Lambda|$ is odd \cite{YangZhang}.

{\sl Lieb's theorem} -- In 1989, Lieb~\cite{Lieb1989} proved that for the attractive Hubbard model,
with $U<0$ (and allowing more generally for site-dependent $U\nd_i<0\ \forall\ i$) that if the total
number of electrons $N$ is even, then the ground state of $\HH$ is unique, and is a total spin singlet,
\ie\ with $S=0$.  A corollary is that if $U>0$ (independent of $i$), $\Lambda$ is bipartite with $|\SA|\ge|\SB|$, and $N = |\Lambda|$ is even, then the ground state of $\HH$ has total spin $S=\half\big(|\SA|-|\SB|\big)$ and degeneracy $2S+1$ (\ie\  spin degeneracy is the only degeneracy).  A convenient example is the so-called Lieb lattice, in which the $\SA$ sublattice is a square lattice and there is a $\SB$ sublattice site at the center of every link (\ie\ the CuO$_2$ lattice of the Emery model
\cite{Emery1987}). Lieb's proof extends the Perron-Frobenius argument deployed in one-dimensional   systems~\cite{LiebMattis1962} by invoking a tool known as ``spin-space reflection positivity". The theorem then entails a hierarchy
of lowest energy levels in different total spin sectors, $E\nd_0(S) < E\nd_0(S+1)$, where $S\in\big\{\half(|\SA|-|\SB|),\ldots,\half|\Lambda|-1\big\}$.  For bipartite lattices, the conditions of Lieb's theorem guarantee a ferromagnetic ground state when there is a sublattice imbalance, \ie\ $|\SA|\ne |\SB|$.  

{\sl Thouless-Nagaoka-Tasaki theorem} -- An extension by Tasaki~\cite{Tasaki1989,Tasaki-book} of the celebrated Nagaoka theorem~\cite{Nagaoka1966}, which in fact was originally proven in a more restricted form by Thouless~\cite{Thouless1965},  states the following: For the Hubbard model with $U=\infty$ and arbitrary non-negative $t\nd_{ij}$, and with $N=|\Lambda|-1$, the ground state has total spin $S=\half N$ and degeneracy $2S+1$.  This establishes that there is a ferromagnetic ground state in the $U\to\infty$ limit when the system is one electron shy of half-filling.  No rigorous extension of these results to finite doped hole density has yet been achieved.

\begin{table}[!t]
\label{LSMOH} 
\begin{center}
\caption
{Possible phases consistent with the LSMOH theorem.
{A check in the first column indicates that the number of electrons per site is not an even integer. ``Unique'' indicates whether a single, non-degenerate ground state on the torus is possible. By ``featureless'' we mean that there is no spontaneous symmetry breaking of any kind (this does not exclude topological order). The last column names a phase that is consistent with the listed properties, if any.}
}
\begin{tabular}{|c|c|c|c|c|c|} \hline
$n\notin 2\MZ$ & unique & gapped & featureless & insulator & example \\ \hline
\gcmk & \gcmk & \gcmk & \gcmk & \gcmk & \tcr{NOT POSSIBLE} \\ \hline
\gcmk & \gcmk & \rxmk & \gcmk & \rxmk & METALLIC \\ \hline
\gcmk & \gcmk & \gcmk & \rxmk & \gcmk & DENSITY WAVE \\ \hline
\gcmk & \gcmk & \rxmk & \gcmk & \gcmk & GAPLESS SPIN LIQUID \\ \hline
\gcmk & \rxmk & \gcmk & \gcmk & \gcmk & GAPPED SPIN LIQUID \\ \hline
\rxmk & \gcmk & \gcmk & \gcmk & \gcmk & BAND INSULATOR \\ \hline
\end{tabular}
\end{center}
\end{table}

{\sl LSMOH theorem} -- The Lieb-Schultz-Mattis-Oshikawa-Hastings (LSMOH) theorem states, in essence, that {\sl when the filling $n$ is not an even integer, a unique, gapped, featureless, insulating ground state is impossible}.  It is 
useful to first  consider the sort of behavior that is ruled out by the theorem but which may pertain if the conditions of the theorem are not met, examples of which are shown in Table \ref{LSMOH}. The Hubbard model in $d=1$ is a band insulator when $n=0$ or $n=2$, but has a Mott phase for $n=1$ that preserves all symmetries and with no adiabatic connection to a band insulator. 
More generally, the theorem applies to models with any number of conserved flavors; in such a case, it states that a featureless, gapped ground state is impossible if the filling of at least one flavor per unit cell is non-integer.

The original argument, due to Lieb, Schultz, and Mattis (LSM)~\cite{LSM1961}, applied to an $N$-site, spin-$S$ $XXZ$ spin chain with periodic boundary conditions and zero total magnetization. LSM showed that if the ground state $\sket{\Psi\nd_0}$ is a state with crystal momentum $K\nd_0$, \ie\ if $\Ht\,\sket{\Psi\nd_0}=e^{iK\nd_0}\sket{\Psi\nd_0}$, where $\Ht$ is the lattice translation operator and we work in units where the lattice spacing is $a\equiv 1$, then the application of the {\it spin twist operator\/},
\begin{equation}
\hat{V}=\exp\!\bigg({{2\pi i}\over N}\sum_{j=1}^N j\,\HS^z_j\bigg)\quad,
\end{equation}
results in a state $\sket{\Psi\nd_1}=\hat{V}\,\sket{\Psi\nd_0}$ with crystal momentum $K\nd_1=K\nd_0+2\pi S$, satisfying $\sexpect{\Psi\nd_1}{\HH\nd_{XXZ}}{\Psi\nd_1}=E\nd_0+\CO(1/N)$. For $\exp(2\pi i S)=-1$, the states are of different crystal momentum and thus orthogonal.  Thus, in the thermodynamic limit any ground state $\sket{\Psi\nd_0}$ must be degenerate (or gapless). The theorem was extended to more general one-dimensional systems in Ref.~\cite{Yamanaka1997}.

{While the LSM argument does not apply in dimensions $d>1$, it was extended by Oshikawa~\cite{Oshikawa2000} and subsequently rigorized by Hastings~\cite{Hastings2004} to $d$-dimensional systems with periodic boundary conditions (\ie\ on a $d$-torus) and a finite excitation gap. The line of reasoning, inspired by Laughlin's argument for quantization of the Hall conductivity in two dimensions~\cite{laughlinflux}, focuses on the consequences of adiabatic $\phi=2\pi$ $\tU(1)$ flux threading through one of the toroidal cycles, followed by a ``pullback" to the original $\tU(1)$ flux state. In the thermodynamic limit, starting with an initial ground state $\sket{\Psi\nd_0(\phi=0)}$, this procedure results in  a low energy state $\sket{\Psi\nd_1(\phi=0)}$ (similar to the LSM construction), which becomes degenerate with the ground state in the thermodynamic limit.}  The analysis can be applied to bosonic systems as well (see, \eg, Ref.~\cite{Sidd2013,Watanabe2015}).

\section{The Hubbard square and its extensions}

The Hubbard square - that is the four site square ``molecule'' of Hubbard sites - is simple enough that its spectrum can be computed analytically~\cite{schumann}, but already complicated enough that it hosts a variety of non-trivial many-body effects that shed considerable light on the more general problem~\cite{ScalapinoTrugman1996}. The symmetry (\ie\ total spin $S$ and transformation properties under the spatial symmetries of the square) of the ground state for the model with nearest-neighbor hopping, $t$, in different ranges of $U/t$ and for different values of the electron number $N$ between 0 and 4, is given in Table~\ref{tab:4site}. Since in the absence of next-nearest-neighbor hopping, $t'=0$, the model is particle-hole symmetric, the analogous results for $4< N\leq 8$ are easily obtained.

The most remarkable feature of this table is that the ground state for $N=4$ and $U>0$ transforms non-trivially under spatial symmetries - it transforms according to the B$_{1g}$ representation of  the D$_4$ spatial symmetries of the square, (\ie\, it transforms like $x^2-y^2$).  It is easy to prove~\cite{fragile} that no non-interacting model on the square can have this property, \ie\ this is an intrinsically new effect of ``strong'' correlations.  It is thus worthwhile to understand its origin.  

\begin{table}
\caption{Character of the ground state of the positive $U$ Hubbard square.}
 \renewcommand{\arraystretch}{1.3}
    \centering
    \begin{tabular}{c|c|c|c|c|c}
    \hline
 $N$ \ \ \  & $S$ \ \ \  & {\rm symmetry} & {\rm range of $U/t$}  & $E_0$ \ {\rm for} \ $U\ll t$ & $E_0$ \ {\rm for} \ $U\gg t$ \\
  \hline
 \hline
 0 & 0  & 1 & {\rm any}\  $U$  & 0&0\\
 \hline
 1 & 1/2  & 1 &   {\rm any} \ $U$ & $-2t$&$-2t$\\
 \hline
 2 & 0  & 1 & {\rm any} \ $U$ & $-4t + U/ 4 -(5/128){U^2}/ t$ & $-2\sqrt{2}t - 4 {t^2}/U$\\
 \hline
\multirow{2}{*} 3 & 1/2  &\   $x,y$\ \  & \  $U < U_{\rm Nag}$ & $-4t + U /2 - (7/128){U^2}/ t$ & ---\\ 
 \cdashline{2-6}
   & 3/2  & 1 & $U_{\rm Nag} < U$ & --- & $-2t$  \\
 \hline
 4 & 0  & $(x^2-y^2)$ &   {\rm any} \ $U$ & $-4t +  {3U}/{4t}- (13/128) {U^2} /t$ & $- {12t^2}/U$\\
 \hline
 \end{tabular}
    \label{tab:4site}
    \end{table}
\renewcommand{\arraystretch}{1}

It is possible to understand this property in the weak coupling limit. We start from considering the single-particle states for $U=0$. Treating the square as a four site ring, all eigenstates may be labeled by a Bloch wavevector $k$.  Provided $|t'| < |t|$, the single particle states are ordered in energy such that the lowest ($k=0$) state has energy $-2t-t'$, the next are the two-fold degenerate ($k=\pm \pi/2$) with energy $+t'$ that transform as $x$ and $y$, and the highest ($k=\pi$) state has energy $2t-t'$ and transforms as $xy$.  Thus, for $U=0$, the ground state with $N=4$ is any state with two electrons (of opposite spin) in the $k=0$ state, and two electrons in either of the first excited states; it is thus six-fold degenerate.  However, they  can be expressed uniquely as states of given symmetry: there is a triplet of $S=1$ states which transforms trivially under spatial transformations (rotations and mirror reflections), and three $S=0$ states, one (with one electron  each in $k=\pm \pi/2$) which transforms trivially under spatial transformations, and two which are superpositions of states with two electrons either in $k=\pi/2$ or $k=-\pi/2$, which transform either as $xy$ or as $x^2-y^2$.  Of these, the one that is adiabatically connected to the $U>0$ ground state is the latter,
\begin{align}
    \ket{\Psi_{x^2-y^2}} = \frac{1}{\sqrt{2}}\left[ c\yd_{\pi/2,\uar}c\yd_{\pi/2,\dar}
    -c\yd_{-\pi/2,\uar}c\yd_{-\pi/2,\dar}\right] c\yd_{0\uar}c\yd_{0\dar}\ket{0}\quad.
\end{align}
Given that there are no degeneracies once symmetry is imposed, it is clearly possible to incorporate the effects of small $U$ by straightforward perturbation theory.  Hund's first rule, which in this context would be expected to be derivable in first order perturbation theory, would imply that the triplet state would be the most likely candidate ground state. Indeed, for any Slater determinant state, the first order energy is $4UN_{\uparrow}N_{\downarrow}$ where $N_\sigma$ is the number of electrons with given spin polarization, and this takes its minimal value of $3U/4$ in a triplet state with three electrons of one polarization and one electron of the opposite polarization.  However, manifestly $\sket{\Psi_{x^2-y^2}}$ is  not a Slater determinant, and indeed it is easy to show that to first order in $U$, it is degenerate with the triplet state.  Computing the second order correction requires a bit of algebra, but the result (given for $t'=0$ in Table~\ref{tab:4site}) is that the singlet state has the lower energy.  

This is a rare example of a case in which Hund's rule is violated.  Note that the cause of this is a subtle quantum effect -- the ground state in the $U\to 0$ limit is a highly entangled state that does not reduce to a single Slater determinant.

The other case that has interesting structure is the $N=3$ square.  For $U=0$, the ground state is four-fold degenerate, but this degeneracy is fixed by symmetry and hence survives to non-zero $U$.  Specifically, the ground state has $S=1/2$ and $k=\pm \pi/2$, \ie\ it transforms under the spatial symmetries as $(x,y)$. However, for large $U$, we know from Nagaoka's theorem that the ground state must have $S=3/2$.  Since parallel spins do not interact in the Hubbard model, this state is the non-interacting (Slater determinant) state with one electron in each of $k=0$ and $\pm \pi/2$;  the resulting state is manifestly invariant under the spatial symmetries.  It requires a little algebra to derive the critical value of $U=U_{\rm Nag}$ that separates the two regimes;  for $t'=0$, $U_{\rm Nag}=4(2+\sqrt{7})\,t \approx 18.6t$.

It is also interesting to consider an ensemble of Hubbard squares.  If they can exchange particles, we can ask the question:  if we have a total of $N$ electrons (with $0< N \leq 8$) and two molecular Hubbard squares on which to place them, what is the lowest energy state?  Not surprisingly, for $N=8$ it is best to put four electrons on each molecule, while for $N=7$ it is best to place three on one molecule and four on the other. However, for $N=6$, the result is more interesting.  We define the pair binding energy,
\begin{align}
\Delta_{\rm p} \equiv 2\,E(3)-E(4) - E(2)\,,
\end{align}
where $E(n)$ is the $n$-electron ground state energy of a molecule.  
$\Delta_{\rm p}$ is the energy difference between placing 3 electrons (one ``doped hole'') on each molecule, or two electrons (one pair of doped holes) on one and four electrons (no holes) on the other. $\Delta_{\rm p}$ 
is negative for $U$ larger than a certain value $U_{\rm p}$, as one might have expected, but it is positive for $U < U_{\rm p}$, \ie\ there is an effective induced attraction between two doped holes.
(For $t'=0$, $U_{\rm p} \approx 4.584 t$.)  

The existence of a range of $U>0$ in which $\Delta_{\rm p}>0$ constitutes the simplest paradigmatic example of a system in which an effective attraction - indeed induced pairing - arises from purely repulsive microscopic interactions.  Ultimately, it is related to the anomalous stability of the state with four electrons on one molecule.  In the weak coupling limit, the pair-binding can be computed perturbatively as $\Delta_{\rm p} =  A\> U^2/t + \ldots$\,,
where $A$ is a function of $t'/t$ with $A=1/32$ for $t' =0$. 
One can also consider the separate contributions of the kinetic energy (hopping term) and the interaction energy (Hubbard term) to $\Delta_{\rm p}$.  In the weak-coupling limit, pair formation is associated with a cost in kinetic energy which is more than compensated by a reduced cost in repulsion, but for $U_{\rm p}> U>U^\star$ (where $U^\star \approx 2.457\,t$ for $t'=0$), $\Delta_{\rm p}$ gets a positive contribution from a lowering of the kinetic energy and a negative contribution (disfavoring pair-binding) from the interactions.  An intrinsically strong coupling mechanism of ``kinetic-energy driven pairing'' is thus in play for $U^\star < U < U_{\rm p}$.

One final lesson from this concerns the preferred symmetry of a ``pairing'' order parameter. 
Consider the operator $\hat{\Phi} = \sum_{i,j} \phi\nd_{i,j}\, c\nd_{i\uparrow}\, c\nd_{j\downarrow}$ that creates the two-electron ground state by acting on the undoped (four electron) ground state. This operator can be viewed as creating a ``Cooper pair'' of doped holes. As first observed by Scalapino and Trugman~\cite{ScalapinoTrugman1996}, the symmetry properties of $\hat{\Phi}$ are 
determined by the {\em difference\/} of the spatial symmetries of two and four electron ground states. For the case tabulated above valid for $|t'|< |t|$, this means that this operator has $d_{x^2-y^2}$ symmetry.
This illustrates the robust preference for $d$-wave superconductivity that we will see is a generic feature of the Hubbard model on the square lattice.

While the Hubbard square is particularly illuminating as it is analytically solvable, it is worth noting that other small clusters - solvable numerically - can exhibit similar behaviors, and in particular regimes pair binding arises even when $U>0$. Examples of this include the Hubbard model on the tetrahedron (in which pair-binding persists for all $U$), cube, and truncated tetrahedron~\cite{whitechakravarty}.

Distinct phases of matter - and in particular spontaneously broken symmetries - do not arise in finite Hubbard clusters.  However, in certain circumstances, the properties of an extended system consisting of weakly coupled clusters can be inferred from the properties of an isolated cluster in a controlled expansion in powers of the (assumed small) intercluster couplings.  Such an analysis was carried through for a 2d ``checkerboard'' array of Hubbard squares in Refs.~\cite{Altman2002,tsai,Yao2007myriad,Baruch2010,drorcheckerboard,scalettarcheckerboard,Karakonstantakis2011}. Here, the complexity of the various regimes found for the isolated square implies the existence of a still richer phase diagram in the $U-x$ plane.  Not surprisingly, in the range of $U$ for which there is pair-binding on the single square, there arises a substantial portion of this phase diagram in which the ground state is a $d$-wave superconductor - although as a consequence of the reconstructed band-structure implied by the four-site unit cell, the line of nodes fails to intersect the Fermi surface and hence the quasiparticle spectrum is fully gapped.  In addition, there is a variety of possibly insulating charge and spin density wave states, and several distinct Fermi liquid phases among which is one with spin-$\frac{3}{2}$ charge $e$ quasiparticles (which is thus not adiabatically connected to any free electrons state) and one with quasi-particles with spin-$\half$, charge $e$, and an additional orbital pseudo-spin $\half$.

\section{One dimension}
\label{1D}
In a tour-de-force of mathematical physics, the 1d Hubbard chain was exactly solved by the Bethe-ansatz method\footnote{The Bethe Ansatz solution allows to calculate the spectrum and the many-body eigenstates of the one-dimensional Hubbard model. Calculating correlation functions is more difficult.}~\cite{Lieb1968}.  For this as well as more general 1d problems, such as the Hubbard model on various relatively narrow width ladders and cylinders, much more is known than for higher dimensional cases.  Particularly powerful numerical methods can be deployed, as discussed in \S~\ref{DMRG} below.  In addition, a weak-coupling renormalization group (RG) - known colloquially as ``$g$-ology'' in reference to the various couplings $g\nd_a$ - leads to a rich set of results involving multiple ``intertwined orders'' in the sense that the interactions that promote charge-density-wave, spin-density-wave, and singlet and triplet superconducting correlations evolve in a complex, interrelated fashion under RG. 

Moreover, at long distances and low energies, the properties of all such 1d systems are describable by a limited set of free boson conformal field theories, \ie\ the physics is that of  a small number of weakly interacting,  linearly dispersing bosonic collective modes.  Distinct quantum phases of matter are generically classified by possible discrete broken symmetries, and by the number of gapless modes (\ie\ the ``central charge'') and the quantum numbers (spin, charge, crystal-momentum) associated with each such mode~\cite{linbalentsfisher,emerymezachar}.   The relation between the microscopic fermionic degrees of freedom and the bosonic modes, called ``bosonization'', can be complicated in specific cases, but is in principle understood in general.  

Other than some of the numerical results, this intellectual block is relatively old and well known - much of it dating from the 1970's - and so will not be reviewed here. For a modern treatment, see ref.~\cite{giamarchi2003quantum}.

\section{Asymptotically exact results in special limits}

\subsection{Weak coupling limit}
\label{weak}
\subsubsection{Background}
In this section, we consider the instabilities of the Hubbard model in $d > 1$ in the weak-coupling limit $U/t \ll 1$.  
The goal is to express the properties of the system starting from the band structure that results from diagonalizing the Hamiltonian with $U=0$.  Corrections to all quantities can be expressed as an asymptotic series in $U/t$.  Clearly, at zero temperature, the radius of convergence of this series is zero, since the  behavior is qualitatively distinct for $U/t \rightarrow 0^-$ and $U/t \rightarrow 0^+$:  When $U/t  \rightarrow 0^-$, BCS mean-field theory is asymptotically exact, and there is a  superconducting instability below a characteristic scale $T_{c,-} \sim \exp{\left[-1/(\rho |U|) \right]}$, where $\rho$ is the density of states at the Fermi energy.  The superconductivity is described by an order parameter $\Delta$, 
which has only one sign on the entire  Fermi surface, so that $\langle \Delta \rangle\nd_\textsf{FS} \simeq \textsf{max}(\Delta)\nd_\textsf{FS}$,
where  $\langle \bullet \rangle\nd_\textsf{FS}$ denotes an average over the Fermi surface. 
Such a superconducting state is termed ``conventional" as it arises in many elemental metals where the pairing mechanism is the electron-phonon coupling.  
When  $U/t  \rightarrow 0^+$, as we now argue, there is also a superconducting instability, but with two  qualitatively different features.  First, the superconductivity itself sets in below a parametrically  lower scale $T_{c,+} \sim \exp{\left[-1/(\alpha \rho^2 U^2) \right]} \ll T_{c,-}$, where $\alpha$ is an order unity constant that depends on the entire band structure of the system.  Second, the superconducting behavior is ``unconventional" in the sense that $\langle \Delta \rangle\nd_\textsf{FS} \ll \textsf{max}(\Delta)\nd_\textsf{FS}$.  

These distinctions imply that the ground states on either side of $U/t=0$ are not adiabatically connected. 
The point $U/t=0$ is a peculiar multi-critical point corresponding to  a free Fermi gas. %there are no fluctuations associated with it. 
These sharply distinct asymptotic behaviors are also  more directly manifest in perturbation theory at finite frequency: a subclass of perturbative corrections represented diagrammatically by ladders in the particle-particle (BCS) channel are logarithmically divergent so long as either time-reversal symmetry or inversion symmetry are present in the normal state. 

In a seminal paper, Kohn and Luttinger (KL)~\cite{KohnLuttinger1965} explored how superconductivity arises from repulsive interactions. They considered a 3d electron gas with weak short-ranged repulsive interactions, for which they identified 
an instability to an unconventional superconducting state with non-zero Cooper pair angular momentum. 
In particular, they emphasized  the role played by  Friedel oscillations associated with the existence of a sharp Fermi surface. 
Below, we provide a modern discussion of the problem in the language of the renormalization group (RG), and show that (in contradistinction to the KL analysis) the structure of the particle-hole susceptibility at all energy scales, not just close to the Fermi level, dertermines the superconducting instabilities~\cite{hirschdwave,emerydwave}.  For a recent reviews on superconductivity from repuslive interactions, consult Refs. ~\cite{scalhubbard,Chubukov2013} and references therein.

\subsubsection{The Fermi liquid fixed point}
We first summarize the basic results of the RG formulation of Landau Fermi liquid theory, pioneered by Polchinski~\cite{polchinski} and by Shankar~\cite{ShankarRG1994}, which will be invoked in the discussion that follows. To keep things simple, we will present the key results for a rotationally-invariant Fermi surface.  The generalization appropriate for anisotropic Fermi surfaces, relevant to crystalline systems, is discussed below. 

Instead of attempting to derive a Fermi liquid from microscopics, the strategy is to proclaim a fixed point of the RG 
and then to analyze its stability.  The fixed point theory is similar to the action $S_0$
{in imaginary time} of a free-Fermi gas in d-dimensions with chemical potential $\mu$ and energy dispersion $E(k)$:
\begin{equation}
S_{0} = \sum_{\alpha=\uparrow,\downarrow}\int {d^dk\, d \omega\over(2 \pi)^{d+1}}\> \bar \psi\nd_{\Bk, \omega, \alpha}\, \big( i \omega -E(\Bk) + \mu \big)\, \psi\nd_{\Bk, \omega, \alpha} \,. 
\end{equation}
We  retain frequency modes $\vert \omega \vert $ less than a UV cutoff $\Lambda\ll E\nd_{\rm F}$ (having ``integrated out" higher energy modes) and thus can linearize the dispersion as $E(\Bk) = \mu + \vF k_{\perp} + \cdots$. We express the measure as  $d^d k = k_{\rm F}^{d-1} dk\nd_{\perp} d\HBk$, where $d\HBk$ is a solid angle element on the Fermi surface,
$k_\perp$ is the direction of momentum perpendicular to the Fermi surface, and $\vF$ is the magnitude of Fermi velocity. Redefining fermion fields as $\psi\nd_{k, \omega} \rightarrow k_{\rm F}^{(d-1)/2}\, \psi\nd_{k,\omega}$\,, 
we arrive at the fixed point action for a Fermi liquid: 
\begin{equation}
S_{FL} = \sum_{\alpha=\uparrow,\downarrow}\int\limits_{-\Lambda}^{\Lambda} \!{d \omega\over 2 \pi}\,{d k_{\perp} d\HBk \over(2 \pi)^d}\>  \bar \psi\nd_{k,\omega,\alpha}\, \big( i \omega - \vF k\nd_{\perp} \big) \,\psi\nd_{k,\omega,\alpha} \,. 
\end{equation}

The stability of the Fermi liquid fixed point is determined by power counting
from which one learns that: (i) a constant shift to the kinetic energy is relevant but harmless, as it amounts to a shift in the chemical potential, and (ii) all higher derivative corrections to the kinetic energy are irrelevant, and hence the kinetic energy is governed by a single parameter, the Fermi velocity $\vF$.  

Next we similarly analyze the role of interactions. A key observation is that generic 4-fermion interactions (and all higher order interactions) are irrelevant at the Fermi liquid fixed point, due to the phase space restrictions imposed by the Pauli principle and the Fermi surface.  This result explains in large part the ubiquity of Fermi liquids in nature (the prime example being liquid helium-3) despite the presence of strong interactions in real systems.  Only under two special kinematic circumstances are interactions important.  First, forward scattering interactions are exactly marginal,  and are incorporated into the Landau parameters.  Second, the dimensionless BCS interaction $V_{\ell}$, where $\ell$ labels the pairing channel\footnote{In a rotationally invariant system, $\ell$ corresponds to the angular momentum channel.  More generally, $\ell$ labels the irreducible representation of a crystallographic point group.}, 
is marginal only at tree-level: one-loop  corrections are logarithmically divergent. 
(This is directly related to the properties of the particle-particle ladders that lead to the Cooper instability for negative $U$ discussed above.) Attractive (repulsive) interactions are marginally relevant (irrelevant).  This is captured by the BCS $\beta$-function, obtained by promoting $\Lambda$ to a running scale $\Lambda = \Lambda_0\,\exp(-t)$, and obtaining the flow of the coupling:
\begin{equation}
\label{bcs_shankar}
\beta = {d V\nd_\ell\over d t} = - V^2_\ell \quad.
\end{equation}
Thus, repulsive  BCS interactions $(V_{\ell} > 0)$ weaken and attractive interactions grow, eventually leading to the BCS instability.

At this point, it may seem that the KL instability is absent since we have just concluded that the metal is stable for repulsive interactions.  However, a system with short-ranged repulsive interactions can have effective attractive interactions in some pairing channel $\ell$, in which case the $\beta$-function {\it in that channel} would indicate a superconducting instability.   In the case of the Hubbard model, the bare repulsive $U$ enters only the s-wave BCS channel and is orthogonal to unconventional pairing channels (\eg\ $d$-wave, $p$-wave, etc.).   But as we integrate out high energy modes, we induce attractive interactions in unconventional pairing channels~\cite{kaganandchubukov}, and the dominant one among these leads to a superconducting state.    

The proper description of such effects involves a two-step renormalization group analysis~\cite{Raghu2010}.  In the first step, we integrate out modes in a weakly interacting metal to obtain a description along the lines of the fixed point theory above.  Then, what were formally repulsive interactions at short-distances may manifest themselves as attractive interactions in certain pairing channels.  In the second step, the RG flow of such couplings determines the superconducting instabilities.

\subsubsection{Two-step Renormalization group for the Hubbard model}
$\\ $
{\bf Mode elimination:}
Lattice electrons governed by the Hubbard model are far from any RG fixed point, and thus any perturbative RG in this regime is meaningless. To overcome this apparent obstacle,  we integrate out the high energy degrees of freedom perturbatively, which is well-controlled in the weak-coupling limit.  This is the first step of the analysis.  Upon doing so, we obtain a low energy effective theory consisting of modes within an energy cutoff $\Lambda\nd_0 \ll E\nd_{\rm F}$\,. Generically, the effective field theory is of the Fermi liquid form described earlier, appropriately generalized to account for crystalline anisotropy.  All salient microscopic details (\eg, lattice symmetry, filling, etc.) are encoded in the shape of the Fermi surface, the magnitude of the Fermi velocity as a function of position on the Fermi surface, and the  marginal perturbations of the Fermi liquid, namely the Landau parameters and BCS couplings.  The choice of the cutoff $\Lambda_0$ is largely arbitrary.  It should be sufficiently small that the quasiparticle dispersion can be linearized, but it cannot be exponentially small in the couplings, where perturbation theory breaks down, as we noted above. Importantly, the final results should not depend on the choice of $\Lambda_0$.  

At the end of the first step, we thus obtain an effective Hamiltonian of Fermi liquid  form $H_{\rm eff} = H_0 + H_{\rm BCS}$
where $H_0$ is the kinetic energy keeping just the linearized dispersion about the Fermi energy and 
\begin{equation}
H\nd_{\rm BCS} = -\half\sum_{k, k'} \Gamma_{\alpha,\beta;\gamma, \delta}(\Bk, \Bk') \,\psi\yd_{\Bk, \alpha}\, \psi\yd_{-\Bk, \beta} \,\psi\nd_{-\Bk', \gamma} \,\psi\nd_{\Bk', \delta} \quad.
\end{equation}
When the system of interest has spatial inversion symmetry (``parity"), the BCS kernel above decouples into  
even and odd parity channels.  Constraints from the Pauli-principle require that without spin-orbit coupling even (odd) parity solutions have spin zero (one):
\begin{eqnarray}
\Gamma_{\alpha,\beta;\gamma,\delta}(\Bk, \Bk') = \Gamma_s(\Bk, \Bk') \left( \delta_{\alpha \gamma}\, \delta_{\beta \delta}
- \delta_{\alpha \gamma}\, \delta_{\beta \delta}
\right) + \Gamma_t(\Bk, \Bk')\left( \delta_{\alpha \gamma} \,\delta_{\beta \delta} + \delta_{\alpha \delta}\, \delta_{\beta \gamma} \right)\quad,
\end{eqnarray}
where the subscripts $s(t)$ denote ``singlet'' (``triplet'').
At weak coupling, these quantities can be expressed as an asymptotic series in $U/t$:
\begin{equation}
    \begin{split}
\Gamma_s(\Bk, \Bk') &= \half U +  \frac{1}{4} U^2 \,\big[ \chi(\Bk + \Bk') + \chi(\Bk - \Bk') \big] + \mathcal O(U^3/t^2) \\
\Gamma_t(\Bk, \Bk') &= \frac{1}{4} U^2\, \big[  \chi(\Bk + \Bk') - \chi(\Bk - \Bk') \big] + \mathcal O(U^3/t^2)\quad.
\end{split}
\label{gammas}
\end{equation}
The quantity $\chi(\Bk)$ is the non-interacting static particle-hole susceptibility, 
where from dimensional analysis it follows that $\chi\sim 1/t$.  
It is important to observe that the strength of the interaction involves both large and small momentum transfers on the Fermi surface and the full structure of the susceptibilities $\chi$ determine the effective interaction, not just the subtle non-analyticities of $\chi$ associated with the sharpness of the Fermi surface.  

{\bf RG flow:}
Having obtained the low energy effective Hamiltonian we now proceed to analyze the RG flow of the BCS couplings.  
We define a dimensionless Hermitian matrix 
$g(\BkF,\BkpF)$,
which is the effective interaction $\Gamma(\BkF,\BkpF)$ defined above, evaluated on the Fermi surface and weighted by the density of states in the neighborhood  of the Fermi points $\BkF,\BkpF$:
\begin{equation}
g(\BkF,\BkpF) ={A\nd_{\rm F}\over \left(2 \pi \right)^d} \>
{\Gamma(\BkF,\BkpF)\over\sqrt{\vF(\BkF)\,\vF(\BkpF)}}\quad, 
\end{equation}
where $A\nd_{\rm F} = \int d \BkF$ is the $(d-1)$-dimensional ``area" of the Fermi surface.
The RG flows are computed by promoting the cutoff $\Lambda_0$ to a running scale $\Lambda \rightarrow \Lambda_0\,\exp(-t)$, and the RG equation obeyed by $g$ is the  convolution
\begin{equation}
{dg(\BkF,\BkpF,t)\over dt} =
-\int {d\BpF\over A\nd_{\rm F}}\> g(\BkF,\BpF,t)\,g(\BpF,\BkpF,t)\quad.
\label{betafunction}
\end{equation}
It is most convenient to analyze the RG flows in the representation where $g$ is diagonal. Distinct eigenstates correspond to, but are not fully specified by an irreducible representation (irrep) of the crystalline point group, where 
\begin{equation}
\int {d\BkpF\over A\nd_{\rm F}}\> g(\BkF,\BkpF) \, \psi\nd_{\ell}(\BkpF) = \lambda\nd_\ell \> \psi\nd_\ell(\BkF)\quad.
\label{eigenvalue}
\end{equation}
Here, $\psi\nd_\ell(\BkF)$ is normalized such that $\int d\BkF \,|\psi\np_{\ell}(\BkF)|^2 = A\nd_{\rm F}$.
Distinct eigenvalues do not mix under RG (within the one-loop approximation), and the BCS $\beta$-function for each $\lambda_{\ell}$ is identical in form to Eq.~(\ref{bcs_shankar}).  

Thus, positive eigenvalues (corresponding to repulsive effective interactions) weaken and negative ones grow.  The first eigenvalue to grow to be of order unity indicates an instability of the Fermi liquid towards the corresponding superconducting state. That the bare Hubbard repulsion can generate effective attractive interactions in unconventional pairing channels is the key result of the two-stage RG analysis.  

\begin{figure}
\includegraphics[width=6.0in]{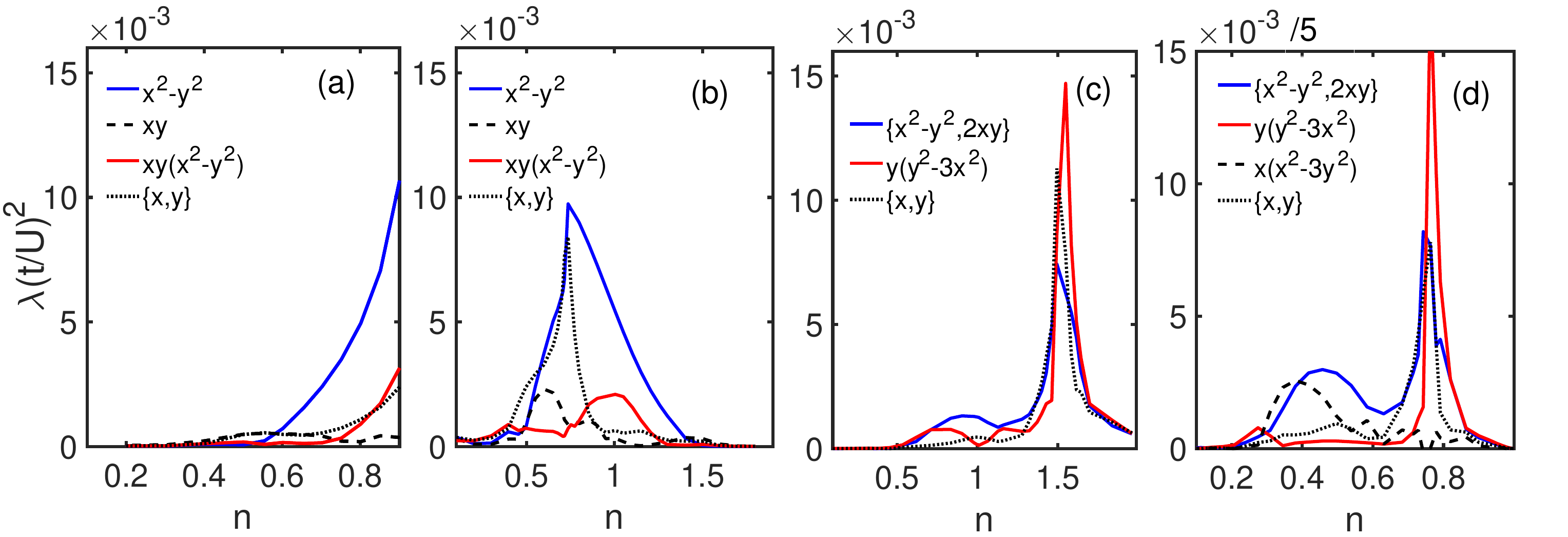}
		\caption{Dominant pairing eigenvalues $\lambda$ (defined in Eq. \ref{eigenvalue}) as a function of $n$ the density per site in the weak-coupling limit of the Hubbard model on a square lattice (a-b) with second-neighbor hopping $t'=0$ (a), $t'=-0.3t$ (b), and (c) a triangular lattice and (d) a honeycomb lattice, both with only nearest-neighbor hopping.   The eigenvalues have been scaled by a factor of 5 in (d) for clarity.    The dominant pairing on a square lattice near half-filling has $d_{x^2-y^2}$ symmetry (labeled $x^2-y^2$) for both $t'=0$ and $t'=-0.3t$.  On the triangular and honeycomb lattices, a two-dimensional  irreducible representation with d-wave symmmetry, labeled $\{x^2-y^2,2xy\}$, is favored when the Fermi surface is electron-like and simply connected.  By contrast, a solution with $f$-wave symmetry, labeled $y(y^2-3x^2)$, sets in near the van Hove filling, and in the regime where there are two hole-like Fermi pockets centered at the zone corners.  The f-wave symmetry corresponds to a nodeless gap function with relative phase of $\pi$  between the two hole pockets.  Solutions with p-wave symmetry, labeled $\{ x,y\}$, also occur in (b-d) in a narrow sliver of energy about the van Hove filling.  This occurs when $n\approx 0.73$ in (b) $n=1.5$ in (c) and $n=0.75$ in (d).  The figures were obtained by discretizing the Fermi surfaces at each density with a grid of 120 points.  Care must be taken in the vicinity of the van Hove fillings.  
		}
		\label{fig:square_lattice}
\end{figure}

In practice, one discretizes the points on the Fermi surface, and solves the resulting discrete eigenvalue problem to find the dominant eigenvalues.  This way, the most important features of the band structures determine the marginal BCS couplings and the dominant instabilities. Fig.~\ref{fig:square_lattice} shows the result for the square lattice Hubbard model with second neighbor hopping $t'=0$ (a) and $t' = -0.3t$ (b).  One sees that the dominant instability of the system in the weak-coupling limit is to a $d$-wave superconductor near half-filling.  In a similar fashion, one can study lattice systems of other geometries~\cite{Hlubina1999,Raghu2010,Deng2015}. Among other things, Ref.~\cite{Deng2015} found a plotting error in Ref.~\cite{Raghu2010}.  Figs.1c and 1d shows results for the triangular and honeycomb lattices respectively with only nearest-neighbor hoppings.  Again, near half-filling, the dominant instability is to d-wave pairing, but in this case this corresponds to a two dimensional irrep, \ie\  a combination of $d_{x^2-y^2}$ and $d_{xy}$ pairing.  Determining which combination of these two components is preferred  in the ordered state is beyond the scope of the RG treatment, but since the effective couplings are weak, this can be addressed within the context of BCS mean-field theory.  As a consequence, one expects the corresponding superconducting state to be the time-reversal symmetry breaking combination generally referred to as $d\pm id$, although a nematic d-wave superconductor (i.e. a real combination of $d_{x^2-y^2}$ and $ d_{xy}$) is also possible.  

{\bf Cutoff-independence:}
A crucial observation made in ref.~\cite{Raghu2010} was that the resulting scale associated with superconductivity is independent of the arbitrary choice of cutoff, $\Lambda_0$.  Indeed, the only characteristic scale, namely the bandwidth $W$ determines the superconducting instability:  $T\nd_{\rm c} \sim W \exp\big(-1\big/\alpha \rho^2 U^2\big)$, as alluded to earlier.  While the details of the proof of cutoff independence can be found in ref.~\cite{Raghu2010}, we provide here some intuition for why this has to be the case.  The main requirement for cutoff independence is that the Fermi liquid fixed point be the only nearby fixed point.  This is certainly true in the weak-coupling limit. 

Let us consider two different mode elimination schemes: (i) eliminating modes above the scale $\Lambda_0$, and (ii) eliminating all modes above $\Lambda_1 < \Lambda_0$. The only assumptions on the cutoffs are that $W \exp(-1/\rho U) \ll \Lambda_1 <  \Lambda_2 \ll U^2/t$.   In the scheme with initial cutoff $\Lambda_1$, we have ``less RG time" $t$ for attractive BCS couplings to grow than the scheme with initial cutoff $\Lambda_0$.  But this discrepancy is precisely compensated by the fact that the effective attractions start off being larger in the scheme with $\Lambda_1$, which is obtained by integrating the $\beta$-function in Eq. (\ref{bcs_shankar}) between $\Lambda_0$ and $\Lambda_1$.  

The presence of other nearby fixed points, or of additional modes (\eg, phonons, magnons, \etc) introduce additional scales and the cutoff independence is then no longer guaranteed.  There is a corollary to this absence of any characteristic scales other than the bandwidth in the weak-coupling limit:  associating a `pairing glue' of a well-developed bosonic fluctuation spectrum, while tempting, is strictly incorrect.  Instead, the electronic fluctuation spectrum itself, at all momentum scales ranging from the lattice scale down to the longest length scales, is responsible for pairing.

\subsubsection{Extensions}
\label{extensions}
In the weak-coupling limit of the Hubbard model in $d>1$ the Fermi liquid state is generically unstable only to superconductivity.  It is typically necessary to treat the model at intermediate or strong coupling to access non-superconducting orders, for instance using the large-$N$ or numerical approaches  discussed below.  To access such orders within the weak-coupling limit, one has to consider non-generic band structures, such as those that produce a perfectly nested Fermi surface or a case in which the Fermi surface crosses a 2d van Hove singularity (vHS).  While these are fine-tuned cases, they bring the interplay between superconductivity and the tendency towards other broken symmetry phases into the weak-coupling regime~\cite{Mark1997}.

Considerable effort has been devoted to such problems.  For instance in the case of Fermi surfaces crossing a 2d vHS, the problem has been studied in straight pertubation theory, first by Dzyaloshinskii~\cite{igor} and Schulz~\cite{schulzvanhove}, who concluded that the most singular tendency was towards superconductivity.  A more sophisticated scaling theory near 2d vHS has remained elusive, as one must contend with the existence of log-squared divergences in the Cooper channel along with log divergences of forward scattering interactions.  The interplay between these effects lead to RG equations that explicitly depend on the energy scale and hence are useless when it comes to constructing asymptotic behavior of correlation functions.  
For a review, see ref.~\cite{Mark1997}. 

As already mentioned, the weak coupling RG in 1d is quite different than for $d> 1$.  However, in special cases, such as for the case of quadratic band-touching in 2d, the RG equations have similar form as in one dimension\footnote{This also applies for quadratic band-touching in 3d with appropriate long-range forces~\cite{Abrikosov1971,Abrikosov1974}.}~\cite{kaiandhong,oskarandkun}.  As in the usual FL, in these cases there are a number of interactions - represented by a set of running coupling constants, $g_a(t)$, that are marginal (dimensionless) by power counting.  To lowest non-trivial order, this leads to a perturbative expression for the RG flows (analogous to that in Eq. \ref{betafunction}) of the form
\begin{align}
   {dg\np_a\over dt} = \Gamma_a^{b,c}\, g\np_b\,g\np_c + \ldots\quad,
    \label{intertwinedbeta}
\end{align}
where summation convention is assumed, and the remaining terms are of third order and higher.  The important point is that the tensor quantity $\Gamma$ which reflects physics of the specific system being studied, generally intertwines the various different interactions in a non-trivial fashion.  

Depending on the specific problem being studied, and the bare values of the interactions, there are two different sorts of behavior solutions of this equation can exhibit.  Under some circumstances, the interactions are all marginal or marginally irrelevant, \ie\ some combinations of $g$'s flow to zero and others do not change under RG (at least to this order).  More usually, there are some set of interactions that are marginally relevant.  In this case, the RG flows carry the system to a ``ray'' along which the interactions continue to grow until, 
no matter how weak the bare interactions, a point is reached at which the perturbative treatment breaks down.  At this point, other methods must be employed to solve the problem.  The possible rays can be identified~\cite{BalentsFisher1996} by looking for possible solutions of the form $g_a(t) = G_a(t^* -t)^{-1}$, where $G_a$ are solutions of the set of quadratic equations
\begin{align}
    G_a  = -\Gamma_a^{b,c}\, G_b\, G_c \quad.
\end{align}
Those couplings for which $G_a \neq 0$ grow strongly toward strong-coupling under the RG flow, while any coupling for which $G_a=0$ remains relatively weak. Such a runaway flow of some of the coupling constants typically signals the opening of an interaction-driven gap, often associated with spontaneous symmetry breaking.

Despite the fact that these results apply (in $d>1$) only for fine-tuned band-structures, they may be useful over some range of intermediate energies and/or temperatures if the fine-tuned conditions are approximately satisfied, especially if one extrapolates the results to intermediate coupling strengths.  We refer the reader to the literature for further discussion of these ideas~\cite{oskarandkun,pujari2016interaction,chubukovspecialbs}. 

\subsection{Strong coupling limit}
\label{strongcoupling}
When the density of electrons per site is $n=1$, the low energy physics in the strong coupling limit of the Hubbard model is, famously, that of the corresponding spin-$\half$ Heisenberg antiferromagnet, with exchange coupling $4\, |t_{ij}|^2/U$\,.
This applies whether the lattice is regular or irregular.  At the same time, the system is a ``Mott'' insulator in the simple sense that the insulating gap 
\begin{align}
\Delta_c= \half U + \CO (t)
\end{align}
is determined by the interaction strength and has no relation to any ordering phenomena that occur below temperature scales of order $J$.  

Needless to say, the physics of quantum antiferromagnets is a rich topic in its own right.  Depending on the lattice structure, the range of the exchange interactions, and the degree of disorder, strong evidence exists for the existence of various forms of antiferromagnetic ordered phases (colinear, coplanar, non-coplanar, commensurate or incommensurate, chiral or non-chiral, etc.), spin-Peierls phases, nematic and spin nematic phases, quantum spin liquid phases of various flavors, spin-glasses, random singlet phases, and surely  others.

For $n< 1$, the low energy physics (\eg, the equilibrium properties the system at temperatures $T \ll U/2$) is governed by a version of the famous $t$-$J$ model
\begin{align}
H_{t-J} = -\sum_{i,j} t\nd_{ij}\, {\hat B}\nd_{ij} + \sum_{i,j} J\nd_{ij}\left( \BS\nd_i\cdot  \BS\nd_j -\frac{1}{4} \,\nhat\nd_i\,\nhat\nd_j\right) - \sum_{i,j,k} K\nd_{ijk}\, {\hat\Delta}\yd_{ij}\,{\hat\Delta}\nd_{jk} + \CO\big(t^3/U^2\big)\quad,
\end{align}
where ${\hat B}\nd_{ij}=c\yd_{i\uar} c\nd_{j\uar} + c\yd_{i\dar} c\nd_{j\dar}$, $\hat \Delta_{ij} = 2^{-1/2}\big(c\nd_{i\uar}c\nd_{j\dar} + c\nd_{j\uar}c\nd_{i\dar}\big)$, and $K_{ijk} = 2t_{ij}t_{jk}/U$, and where it is understood that $H_{t-J}$ operates in a restricted Hilbert space in which no site is doubly occupied. 
This model is often studied in its own right (typically neglecting the term proportional to $K$) and, so long as $J_{ij} < |t_{ij}|$, the results are often taken as representative of results for the Hubbard model in some physically reasonable range of $U_{\rm eff} \sim 4t^2/J$. However, strictly speaking, the mapping to the $t-J$ model is valid only in the asymptotic limit $J_{ij}/t_{ij} \to 0$.

Thus, truly in the strong coupling limit, one should begin with the solution of $U\to \infty$ problem, \ie\ the $t-J$ model with $J_{ij}=K_{ijk}=0$, and only keep corrections due to non-zero $J$ and $K$ as small perturbations. The Hubbard model was originally introduced to study itinerant ferromagnetism, based on the fact that this occurs in Hartree-Fock approximation so long as the ``Stoner'' criterion is satisfied, $U \rho(E\nd_{\rm F}) > 1$. 
However, 
more accurate numerical studies have found that
ferromagnetism in the Hubbard model on a bipartite lattice with $|\SA|=|\SB|$
seems to
require $U\rho(E\nd_{\rm F})\gg 1$.  To elicit ferromagnetism for $U/t=\CO(1)$,
it appears necessary to introduce frustration in the form of further neighbor same-sublattice hoppings, or to consider the model on non-bipartite lattices~\cite{Ulmke1998} or on line graphs~\cite{Mielke1991,Mielke1992} at densities away from half-filling.

Provided the kinetic energy satisfies the Perron-Frobenius condition ($t_{ij} \geq 0$ for all $i$  and $j$), the problem with a single hole in a finite size lattice is governed by Nagaoka's theorem (see \S~\ref{theorems}), \ie\ the ground state is a fully polarized ferromagnetic state.  Clearly, the limit $U\to \infty$ and $N\to \infty$ do not commute.  However, for large but finite $U$, the solution in the thermodynamic limit can easily seen to be a ``Nagoaka polaron.''  Its nature can be understood from a simple variational argument:  Consider a state in which a region of radius $R$ has spins aligned ferromagnetically, while the rest of the system is in its $n=1$ antiferromagnetic ground state.  The variational energy of such a state is
\begin{align}
E_{\rm{Nag}} = -E_0 + \bar tR^{-2} +  \bar J R^d  \quad,
\end{align}    
where $E_0 = \sum_j t_{ij}$, and $\bar t$ and $\bar J$ are averages of $t_{ij}$ and $J_{ij}$ that depend on the details of the lattice structure and the assumed shape of the polaron;  here, the second term is the kinetic energy cost of confining the hole in a region of size $R$ and the third term is the cost in exchange energy of making a ferromagnetic bubble.  Minimizing this expression with respect to $R$, we obtain an expression for the size and energy of the Nagaoka polaron:
\begin{align}
E\nd_{\rm Nag} = -E_0+2\bar tR^{-2}  \ \ {\rm with} \ \ R\nd_{\rm Nag}= \big(2\bar t/d\bar J\big)^{1/(d+2)}\quad.
\end{align}

It is a straightforward exercise to go from the single polaron to the problem with a small but finite hole density.  The polaron should strictly be viewed as a new sort of quasi-particle - one with charge $e$ and spin $S \sim R_{\rm Nag}^d$. Two such particles have a short-range effective attraction of order $\bar tR_{\rm Nag}^{-2}$; if two are placed adjacent to each other, the hole associated with each polaron can now delocalize over twice as large a ferromagnetic region.  Thus, there is a tendency for polarons to agglomerate, \ie\ for the system to phase separate.  Opposing this is the effective Fermi pressure of the polaron gas -- however, because the polaron is large,  its effective mass is large,\footnote{Relative to the band mass, the Nagaoka polaron can readily be seen to have  has an effective mass enhancement of order $(\bar t/\bar J) R_{\rm Nag}^{d-1}$. This is since moving the polaron involves flipping spins along its entire surface.}
so the Fermi pressure is negligible.  Thus, at $T=0$ and for low density of doped holes, $1 > n > n_c$, one expects macroscopic two-phase coexistence between an undoped antiferromagnetic phase with $n=1$ and a fully polarized ferromagnetic phase with $n=n_c$ where $n_c \sim R_{\rm{Nag}}^{-d}$.  Correspondingly, for a range of densities $n> n_c$ but not too small, this line of reasoning leads one to expect a half-metallic ferromagnetic phase, \ie\ a state with all the spins parallel to one another.  

The stability of the fully spin-polarized state (known as a half-metallic ferromagnet) at finite doped hole density has not been rigorously established.  Indeed, it has been shown that even for $U=\infty$, the fully polarized state is unstable beyond a (typically substantial) critical doped hole density~\cite{krishnamurthy}. However, exact diagonalization~\cite{ekl} and DMRG studies~\cite{liuferro} of the model on a square lattice (2d) provide strong corroborating evidence that the half-metallic ferromagnetic phase is stable for  $U=\infty$ in the range of density $1> n\gtrsim 0.8$ and that two-phase coexistence occurs at large but finite $U$ in a range of density\footnote{It is a peculiarity of the Hubbard model that there is no interaction between electrons with the same spin -- thus, for any value of $n$, all energy eigenstates with maximal total spin are simply Slater determinants.  In particular the half-metallic ferromagnetic ground state is always an eigenstate of the Hubbard Hamiltonian - the only question is under what circumstances it is the ground state.} $1 > n>n_c \sim (t/U)^{1/2}$.
However, there is also indication that ferromagnetic phases arise only when $U$ is extremely large, $U/t \gtrsim 100$.  None-the-less, the existence of ferromagnetic phases at very large $U$ implies that, strictly speaking, the non-ferromagnetic states generally seen (and expected) at intermediate values of $U$ cannot be approached from a strong-coupling perspective. 

\subsection{Dilute limit}
The dilute limit of the Hubbard model is defined by fixing $U/t$ and letting $n\rightarrow0$. In this limit, strong arguments have been put forward that in $d=2$ and $3$ and for positive $U$, the system forms a Fermi liquid (which may have a superconducting instability at very low temperatures, to be discussed below). 

Consider $d=3$ first. Using the filled Fermi sea as a trial state,
the kinetic energy per particle scales as $\mathcal{E}_{\rm KE}\sim t\, n^{2/3}$
(where we have used the effective mass approximation near the band
bottom), whereas the typical interaction energy satisfies $\mathcal{E}_{\rm PE}\sim Un$.
Thus, in the limit $n\rightarrow0$, $\mathcal{E}_{\rm K}\gg\mathcal{E}_{\rm PE}$.
One may therefore expect the system's properties to be calculable
in an expansion in the small parameter $(U/t)\cdot n^{1/3}$ (which plays a similar role to that of $r_s$ in a uniform electron gas). As explained
in Refs.~\cite{abrikosov1958,Engelbrecht1992,galitskii1958}, the proper small parameter
is actually
$\kF  a\nd_{\rm s}$, 
where $\kF\propto n^{1/3}$ and $a\nd_{\rm s}$ is the scattering length. For small $U/t$, the scattering length satisfies
$a_{\rm s} \sim a\,(U/t)$ (where $a$ is a lattice constant). 

In $d=2$, the situation is more subtle, since
the above argument gives that $\mathcal{E}_{\rm KE}/\mathcal{E}_{\rm PE}$ is
density independent. However, a more careful treatment (outlined below)
shows that in this case the two-particle scattering amplitude near
the Fermi energy, propotional to $1/\ln(1/n)$, serves as an emergent
small parameter~\cite{Engelbrecht1992}. Thus, in both cases, a systematic
expansion starting from the Fermi gas is possible, resulting in a
Fermi liquid whose Landau parameters are parametrically small in the
$n\rightarrow 0$ limit. 

\begin{figure}

\includegraphics[width=0.85\textwidth]{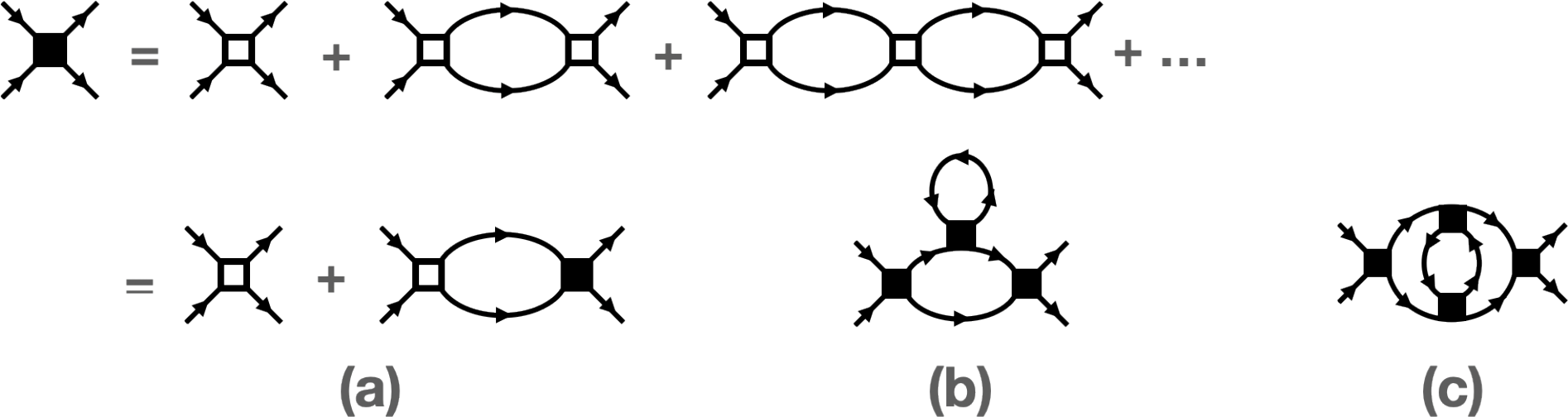}\caption{(a) Ladder series for the two-particle scattering amplitude in the
dilute limit. The empty square represents the bare interaction $U$,
and the filled square is the effective interaction $U_{\text{eff}}$
(also known as the T matrix). (b,c) Examples of diagrams which are
not part of the ladder series. These diagrams are suppressed comapred
to those shown in panel in the $n\rightarrow0$ limit (a) (see text).}
\label{fig:ladder}
\end{figure}

To arrive at the expansion appropriate for the dilute limit, we examine
the diagrammatic representation of the two-particle vertex function.
The dominant terms are the ones that form the ladder series [Fig.~\ref{fig:ladder}(a)]. Summing the ladder series gives the effective
two-particle interaction (also known as the $T$-matrix):

\begin{equation}
U_{\text{eff}}(\bm{q},\epsilon)={U\over 1+\Gamma\nd_0(\bm{q},\epsilon)\,U}\quad,
\end{equation}
where $\Gamma\nd_0(\bm{q},\epsilon)=\int\frac{d^{d}k}{\left(2\pi\right)^{d}}\,(\varepsilon_{\bm{k}}+\varepsilon_{-\bm{k}+\bm{q}}-i\epsilon)^{-1}$.
In the dilute limit, we are interested in the interaction at momenta and energies of the order of the Fermi momentum and Fermi
energy, respectively. In $d=3$, we can safely take $\left|\bm{q}\right|\rightarrow0$
and $\epsilon\rightarrow0$, obtaining $\Gamma\nd_0\sim 1/t$. 
Terms that are not part of the ladder series, such as those shown
in Fig. \ref{fig:ladder}(b,c), are suppressed by powers of 
$\rho U_{\text{eff}}\sim \kF a\nd_{\rm s}$, and are thus small. The effective interaction near the
Fermi energy is thus finite in the dilute limit. The Fermi liquid
parameters, such as the Landau function and the effective mass correction,
are all of the order of $\kF a\nd_{\rm s}$ (since they are proportional to
$\nu_{0}\propto \kF $). The smallness of the Fermi liquid parameters
ensures the self-consistency of the expansion. 

In $d=2$, $\Gamma\nd_0$ diverges logarithmically in the limit $\left|\bm{q}\right|\rightarrow0$,
$\epsilon\rightarrow0$. This divergence is cut off by setting $\left|\bm{q}\right|\sim \kF $,
$\epsilon\sim\mu$, where $\mu$ is the chemical potential. In the
dilute limit, such that $(U/t)\ln(1/n)\gg 1$, this gives $U_{\text{eff}}\sim t/\ln(t/\mu)\sim t/\ln(1/n)$.
Hence, terms that are not part of the ladder series are suppressed
by a factor $\rho U_{\text{eff}}\sim1/\ln(1/n)\ll1$, and can be
neglected~\cite{Engelbrecht1992}. The Fermi liquid parameters are small in
proportion to $1/\ln(1/n)$. 

In both $d=2$ and $d=3$, the system is thus described as a weakly
interacting Fermi liquid in the dilute limit, independently of the
original Hubbard interaction. The Fermi liquid state can then be treated
as discussed in \S~\ref{weak}. In $d=3$, and in the presence of time reversal
or inversion symmetries, one expects an instability towards a triplet
superconducting state whose critical temperature $T\nd_{\rm c}$ scales as
\begin{equation}
E\nd_\text{F}\, e^{-1/(\rho U_{\text{eff}})^{2}}
\sim  t\, n^{2/3}\,\exp\left\{-\left(1+\frac{t}{U}\right)^{2}n^{-2/3}\right\}\quad,
\end{equation}
where we have omitted dimensionless numbers of the order of unity
in the exponents. The $d=2$ case requires additional care, since the Lindhard susceptibility is nearly momentum
independent. Hence, no effective attraction is generated at second
order in $U_{\text{eff}}$. A calculation of the third order terms,
performed in ref.~\cite{Chubukov1993}, gives $T\nd_{\rm c}\sim
E\nd_{\rm F}\,\exp\!\big[\!-1/(\rho U_{\text{eff}})^3\big]\sim
t\,n\exp\!\big[\!-1/\ln^{3}(1/n)\big]$.
Note that in both $d=2$ and $3$, $T\nd_{\rm c}\ll E\nd_{\rm F}$\,. 

\subsection{Large-$N$ generalization and the Hartree Fock approximation}\label{largeN}
\subsubsection{A large $N$ limit}
While quantum many-body models are generally insoluble except in very special cases, by extending the global symmetry of the model from $\tSU(2)$ to a larger group such as $\tSU(N)$ or $\tSp(N)$, a systematic expansion in powers of $1/N$ about the $N\to\infty$ limit can often be derived~\cite{Affleck1988,AA88,Marston1989,ReadSachdev1991,Hewson-book,Assa-book}. 
Here, we discuss a large-$N$ generalization of the Hubbard model~\cite{fradkin2013field} that allows controlled access to a variety of intermediate-coupling spin and charge ordered phases. Consider the Hamiltonian
\begin{equation}
H=-\sum_\ij\sum_{\alpha=\pm}\sum_{m=1}^N c\yd_{i\alpha m}
\big(t\nd_{ij} + \mu\,\delta\nd_{ij}\big) c\nd_{j\alpha m} + {1\over N} \sum_i \big(\half V n^2_i - \half J\BS^{\,2}_i + \half K \RPsi\yd_i\RPsi\nd_i\big)\quad,
\label{eq:HN}
\end{equation}
where $i$ labels the lattice sites, $\alpha=\pm$ the spin polarization, and
$m\in\{1,\ldots,N\}$ is a `flavor' index.  The coupling constants $V$, $J$, and $K$ are all non-negative. The local density $n\nd_i$, spin $\BS\nd_i$,
and superconducting order parameter $\RPsi\nd_i$ are given by
\begin{equation}
n\nd_i=\sum_{\alpha,m} c\yd_{i\alpha m} c\nd_{i\alpha m} \quad,\quad
\BS\nd_i={1\over 2}\!\sum_{\alpha,\beta,m} c\yd_{i\alpha m} \,
\Bsigma\nd_{\alpha\beta}\,c\nd_{i\beta m} \quad,\quad
\RPsi_i=\sum_{\alpha,\beta,m} c\nd_{i\alpha m}\,\eps\nd_{\alpha\beta}\, c\nd_{i\beta m} \quad,
\end{equation}
where $\eps\nd_{\alpha\beta}=i\sigma^y_{\alpha\beta}$. Consider a global transformation of the fermion fields
\begin{equation}
c\nd_\ism \to {\widetilde c}\nd_\ism = \CR\nd_{mm'}\,\CU\nd_\ssp\,c\nd_{i\sigma' m'}\quad.
\end{equation}
Suppressing the site index $i$, we write ${\tilde c}=\CR\,\CU\, c$, where $\CR$ acts on flavor indices
and $\CU$ acts on spin indices.  Thus,
\begin{equation}
\begin{split}
{\widetilde n} &= c\yd_{\sigma m}\, (\CR\yd \CR)\nd_{mm'}\, (\CU\yd \CU)\nd_\ssp\, 
c\nd_{\sigma' m'} \\
{\widetilde S^a} &= \half\,c\yd_{\sigma m} \, (\CR\yd \CR)\nd_{mm'}\, 
(\CU\yd \sigma^a\, \CU)\nd_\ssp\,c\nd_{\sigma' m'}\\
{\widetilde \RPsi} &= (\CR^\tt \CR)\nd_{nn'}\, (\CU^\tt \eps\,\CU)\nd_{\sigma\sigma'}\,
c\nd_{\sigma n\np}\,c\nd_{\sigma' n'}\quad.
\end{split}
\end{equation}
Thus if $\CU\in\tSU(2)$ and $\CR\in\tO(N)$, we have ${\widetilde n}=n$, ${\widetilde\RPsi}=\RPsi$, 
and furthermore ${\widetilde S}^a = M\nd_{ab}\,S^b$, where $M\nd_{ab}=\half\,\Tr\, \big(\CU\yd\,\sigma^a\,\CU\nd\,\sigma^b\big)\in\tSO(3)$.  Hence, ${\widetilde H}=H$, 
\ie\  the Hamiltonian H possesses a global $\tU(2)\times\tO(N)$ symmetry (including the $\tU(1)$ charge conservation). {In the case $N=1$, the system reduces to the usual Hubbard model with $U = V + \frac{3}{4}J + 2K$.}

To elicit the large-$N$ theory, we employ three Hubbard-Stratonovich transformations to
decouple each of the three terms quartic in the fermion operators.  The resulting dimensionless
action is then
\begin{equation}
\begin{split}
A&=\int\limits_0^{\beta} d\tau\>\sum_i \bigg( {N\over 2V} \phi_i^2 +
{N\over 2J}\,\Bchi_i^2 + {N\over 2K} \,|\Delta\nd_i|^2\bigg)+
\sum_{i,j\atop\sigma,m}{\bar c}^{\vphantom{\dagger}}_{i\sigma m}\,
\big(\pz\nd_\tau - t_{ij} -\mu \delta_{ij}\big)\,c\nd_{j\sigma m} \\
&\hskip 0.15in - {1\over 2}\sum_{i,m}\sum_{\alpha,\beta} 
\begin{pmatrix} \cbar\nd_{i\alpha m} & c\nd_{i\alpha m}\end{pmatrix}
\begin{pmatrix} i\phi\nd_i\,\delta\nd_\ttp + \half \Bchi\nd_i\cdot\Bsigma\nd_\ttp &
i\Delta\nd_i\,\eps\nd_\ttp \\ i \Dbar_i\,\eps\nd_\ttp &
-i\phi\nd_i\,\delta\nd_\ttp - \half \Bchi\nd_i\cdot\Bsigma\nd_\ttp\end{pmatrix}
\begin{pmatrix} c\nd_{i\beta m} \\ \cbar\nd_{i\beta m} \end{pmatrix}\ .
\end{split}
\label{eq:A}
\end{equation}
Here $\beta=1/k\nd_{\rm B} T$ where $T$ is the temperature, $\tau$ is imaginary time, and
$\big\{\phi\nd_i,\Bchi\nd_i,\Rep\,\Delta\nd_i,\Imp\,\Delta\nd_i\big\}$ are the six time-dependent Hubbard-Stratonovich
fields.

We may now formally integrate out the fermions, obtaining the following effective action:
\begin{equation}
A_{{\rm eff}}=\int\limits_0^\beta d\tau\>\sum_{i}\left({N\over 2V}\,\phi_{i}^{2}+{N\over 2J}\,\bm{\chi}_{i}^{2}+{N\over 2K}\,\left|\Delta_{i}\right|^{2}\right)-\frac{1}{2}N\,{\rm Tr}\log\mathcal{G}^{-1}[\{\phi_i\},\{\bm{\chi}_i\},\{\Delta_i\}]\quad.
\end{equation}
Here, $\mathcal{G}^{-1} = \mathcal{G}^{-1}_0 - \mathcal{M}[\{\phi_i\},\{\bm{\chi}_i\},\{\Delta_i\}]$, where 
\begin{equation}
    \mathcal{G}_{0}^{-1}=\left(\begin{array}{cc}
G_{0}^{-1} & 0\\
0 & -\left(G_{0}^{-1}\right)^{T}
\end{array}\right)\quad.
\end{equation}
The inverse of the free Green's function is given by $G_0^{-1} = -\partial_\tau + t_{ij} + \mu\delta_{ij}$, and $\mathcal{M}$ is the matrix that appears in the second line of Eq.~(\ref{eq:A}). 

Crucially, the effective action is proportional to $N$. Hence, in the large $N$ limit, the partition function is dominated by the lowest-action saddle point of $A_{\rm{eff}}$. Fluctuations around the saddle point are suppressed by powers of $N^{-1}$. We seek a saddle point characterized by time-independent, but possibly site-dependent fields $\phi\nd_i$ and $\Bchi\nd_i$. 
Since we have assumed $K\geq 0$, it is straightforward to see that $\Delta_i =0$ at any such saddle point.
Substituting  $\widetilde{\phi}\nd_i\equiv i\phi\nd_i$ and differentiating the effective action with respect to $\widetilde{\phi}\nd_i$ and $\Bchi\nd_i$ yields:
\begin{equation}
\begin{split}
\widetilde{\phi}\nd_i	&=-V\sum_{\alpha,m}\big\langle\cbar\nd_{i\alpha m}\,c\nd_{i\alpha m}\big\rangle\nd_{\widetilde{\phi}\nd_{i},\Bchi\nd_{i}} \\
\Bchi\nd_{i}	&=\half J\!\sum_{\alpha,\beta,m}\big\langle\cbar\nd_{i\alpha m}\,\Bsigma\nd_{\alpha\beta}\,c\nd_{i\beta m}\big\rangle_{\widetilde{\phi}\nd_i,\Bchi\nd_i}\quad.  
\label{eq:HF}
\end{split}
\end{equation}
Here, the expectation value is taken with respect to the quadratic action (\ref{eq:A}) with $\{\widetilde{\phi}_{i}\},\{\bm{\chi}_{i}\}$ set to their saddle point values. Eqs.~(\ref{eq:HF}) are the self-consistent Hartree-Fock (HF) equations describing possible spin and charge ordered states. They are identical to the HF equations of the one-band Hubbard model with $2V = \half J = U$, $K=0$. Thus, the Hartree-Fock approximation becomes asymptotically exact in the large-$N$ generalization of the Hubbard model given by Eq.~(\ref{eq:HN}).\footnote{Interestingly, setting $N=1$, this choice of parameters corresponds to a Hubbard model with an interaction strength $V + \frac{3}{4}J + K = 2U$. This discrepancy between the Hartree-Fock equations for the Hubbard model and the saddle point equations obtained by decoupling the interaction in an SU(2) symmetric way was noted in Ref.~\cite{schulz1995functional}.} 

\subsubsection{Some Hartree-Fock results}
\label{hartreefock}
There have been numerous Hartree-Fock studies of the Hubbard model.  A priori - especially when dealing with forms of order that only arise when $U$ exceeds a finite critical value - there is no obvious small parameter that justifies these solutions.  However, with the large $N$ limit as justification, it is worth summarizing at least some of the Hartree-Fock results that have been obtained in this way.

In the case of a half-filled band, $n=1$, with $t' =0$, the Fermi surface is perfectly nested and the HF ground state is a N{\'e}el antiferromagnetic insulator for all $U>0$.  In particular, for $d>2$ and small $U$, the sublattice magnetization, $m$, and the quasi-particle gap, $\Delta\nd_{\rm AF}$\,, depend on $U$ as $m \sim \Delta\nd_{\rm AF} \sim \exp(- 1/\rho U)$,  
while for $d=2$ the expressions are slightly more complicated, with $m \sim \Delta\nd_{\rm AF} \sim \exp\!\big(\!-\sqrt{8\pi^2 t/U}\big)$, because of the logarithmically divergent density of states associated with the van-Hove points.
For small but non-zero $t'$, the system remains metallic for $U < U\nd_{\rm c}$  but  is antiferromagnetic for larger $U$, where
 $U_{\rm c} \sim t/\ln|t/t^\prime|$ for $d>2$ and $U_{\rm c} \sim t/\ln^2|t/t^\prime|$ in $d=2$.
The details of the metal-insulator transition as a function of $U$ has not been exhaustively studied, and may vary depending on dimensionality, the value of $t^\prime$, and other details.  The simplest cases involve either a direct first order transition from a featureless metal for $U<U\nd_{\rm c}$ to  an AF insulator for $U>U\nd_{\rm c}$, or a sequence of two transitions, the first (typically continuous) from a featureless metal for $U<U\nd_{\rm c1}$ to an antiferromagnetic metal phase - \ie\ with  $m$ small enough that a portion of the Fermi surface remains ungapped - followed by a (typically first order) transition to the AF insulator at $U> U\nd_{\rm c2}>U\nd_{\rm c1}$.

It is also interesting to consider the evolution of the HF ground state with the introduction of a dilute concentration of doped holes,  $x\equiv 1-n\ll 1$, starting from the AF insulator.  One possible solution of the HF equations is simply a doped version of the N{\'e}el state - here the energy cost per doped hole is $\Delta\nd_{\rm AF}\left\{1 + {\cal O}(x^{2/d})\right\}$.  This is typically never the lowest energy HF solution~\cite{SKLosAlamos};  a better solution is one in which a portion of the sample is undoped AF and another is a non-magnetic metal with a finite concentration $x_c$ of doped holes, where  $x_c \sim \rho(E\nd_{\rm F})\, \Delta\nd_{\rm AF}$.  For small $U$ in $d>2$, it was shown in  ref.~\cite{SKLosAlamos} that $x_c = \sqrt{2}\, \rho(E\nd_{\rm F})\, \Delta\nd_{\rm AF}$ and that the energy per doped hole is $2^{-1/2}\Delta\nd_{\rm AF}\left\{ 1+ {\cal O}(x_c^{2/d})\right\}$, \ie\ it has lower energy than the doped N{\'e}el state.  The same analysis applies for small $t^\prime$ in $d=2$, and even for $t^\prime=0$ with  with slight complications due to the divergent density of states.  
However, more interesting insulating ``stripe'' states were found~\cite{zaanenstripes,shultzstripes1,machidastripes1} to have still lower energies - at least in $d=2$ with $t^\prime=0$. Here for small $x$, the system forms a unidirectional spin-density wave (SDW) state in which the doped holes are localized on anti-phase domain walls a distance $W$ apart, resulting in a new periodicity of the spin-order $\lambda = 2W$.  Moreover, the domain-wall spacing is such that there are an even-integer number of electrons per unit cell, \ie\ $\lambda =1/x$, and the system remains insulating.  From numerics in 2d with $t^\prime=0$, the energy per doped hole of the striped phase was estimated~\cite{shultzstripes2} to be approximately $0.66\,\Delta\nd_{\rm AF}$, some 7\% less than that of the phase separated state.

\section{Numerical results}

\subsection{DMRG results for ladders and cylinders}
\label{DMRG}
The density-matrix-renormalization group (DMRG) approach~\cite{White1992DMRGs,schollwoek2} has proven to be extremely useful in obtaining ground state correlations of Hubbard cylinders and ladders.  As the computational effort grows roughly linearly with the length of the system, $L$, but exponentially with the number of legs, $W$, these results are largely confined to rather small $W$.  However, for these systems, it provides an incomparably versatile approach to the intermediate coupling problem, without relying on any artificial limiting procedure. 
\subsubsection{The utility of DMRG }
DMRG is now understood to be an extremely clever variational approach~\cite{ostlund,schollwoek2} to studying the ground state properties of arbitrary interacting electron models.  It uses a class of variational states known as matrix-product states - which as the name suggests are parameterized by the elements of a matrix associated with each {site}. The larger the dimension of the matrices (known as the ``bond dimension''), the better the approximate ground state obtained.  For any finite size system, DMRG calculations in principle converge to the exact result for large enough bond dimension, $B_d$.  However, as the calculations are more demanding the larger $B_d$, not all published results can be taken at face value. This is reflected in the unfortunate fact that there are examples in the literature of DMRG studies that have reached contradictory conclusions concerning the character of the ground state phase of certain model problems.  Moreover, different sorts of results require different levels of care~\cite{TroyerDMRGcritique}, as we elaborate below.
\begin{itemize}
\item
Generally, the bond dimension required to achieve a given accuracy increases exponentially with the degree of entanglement of the phases being explored.  Thus, DMRG results converge more easily in systems in which the central charge (the number of gapless modes) is small and ones in which all correlation lengths are short. Conversely, systems with many gapless modes and/or long correlation lengths may need a very large $B_d$ to converge.
\item	
Even within a given phase, the reliability of the DMRG results for fixed $B_d$  depends, to some extent, on what questions are  asked. The nature of the short-range correlations  – \ie\ what sort of “local order’’ arises – is less sensitive to subtle aspects of the entanglement, and so can be extracted reliably even in calculations with relatively modest values of $B_d$, while issues concerning long distance correlations – especially power-law falloff of correlations associated with gapless modes – are much more strongly dependent on $B_d$.
\end{itemize}

In quoting results of DMRG calculations, we have restricted ourselves to either discussing results that pertain to short-range correlations (about which there typically is no disagreement), or to results in which the $B_d$ dependence has been seriously investigated and the extrapolation to $B_d\to \infty$ appears convincing. 
There are also generalizations of DMRG, such as projected entangled pair states (PEPS)~\cite{PEPS06}, 
which uses tensorial generalizations of the matrix product states of DMRG. Because these methods have not been as thoroughly tested and benchmarked as has DMRG, we have largely omitted results obtained from these approaches.  
DMRG methods have recently been extended and are beginning to be applied to study dynamical properties of ladders and cylinders~\cite{WhitetDMRG,feiguntDMRG}. This is a very promising new direction, but one that is still in its infancy.  Again, taking a conservative stance, we have mostly not reviewed these results either. (See, however, Ref. \cite{qin2021hubbard}.)

The bulk of the DMRG studies have been carried out on ladder or cylinder versions of Hubbard model for ``intermediate'' values of $U$ of order the band-width, $8t \leq U \leq 12t$, and for ranges of electrons per site $n$ in the range $1.3 \geq n \geq 0.7$, corresponding to a concentration of ``doped holes'' or ``doped electrons'' $0\leq x \leq 0.3$.    Effects of differing band-structure have been studied largely by considering a range of first and second neighbor hopping, $t^\prime/t$. 
Many studied have treated the $t-J$ instead of the Hubbard model, because the results for the former tend to have better convergence properties. Results so obtained are generally interpreted as representative of the solutions of a corresponding Hubbard model with $U/t \approx (4t/J)$ and a somewhat renormalized value of $t^\prime/t$. 

We now summarize some of the salient results that have been obtained from these studies.  We distinguish three types of inferences that can be drawn: 
\begin{itemize}
\item[{\bf 1)}] Most directly, from an analysis of the short-distance behavior of various ground state correlation functions, it is relatively straightforward to determine what sort of ordering tendencies are strong for particular ranges of parameters.  
We will illustrate such orders by describing the form of broken symmetry (long-range order) that would result were these short-distance correlations extended to long distances.  Since the Hubbard model is not a realistic model of any actual material, it may be that identifying  ordering {\it tendencies} reveals the most model independent - and hence physically significant - features of the interesting strong correlations physics.  
\item[{\bf 2)}]  Extrapolating the result to the limit of infinitely long cylinders and ladders, we can identify distinct phases of matter based on the asymptotic long-distance correlations.  As discussed in \S \ref{1D}, this means identifying any discrete broken symmetries, and the number and character of distinct gapless charge and spin modes;  for instance, a Luther-Emery liquid has a single gapless charge mode (central charge $c=1$), and power-law (QLRO) CDW and superconducting correlations.  
\item[{\bf 3)}] Most speculatively, we can introduce conjectures concerning the extrapolation of the DMRG results to the 2d ($W\to\infty$) limit. This can only be done with confidence when the results are weakly $W$ dependent even for relatively small $W$.
\end{itemize}

\subsubsection{The square lattice}  \

{\bf Undoped ladders}:
For the most part, studies of the undoped model ($x=0$) have been carried out in the regime in which $U$ is (assumed) large enough that there is a substantial charge gap, $\Delta_c\gtrsim t$, so the Hubbard model is  equivalent to a spin $1/2$ Heisenberg model with first and second neighbor exchange couplings, $J^\prime/J = (t^\prime/t)^2$.  For even $W$, this necessarily results in  a fully gapped state~\cite{haldaneconj,ChakravartyPRL}, \ie, with a non-zero spin-gap, $\Delta_{\rm s}$.   

\begin{figure}[t]
	\begin{center}
		\includegraphics[scale=0.4]{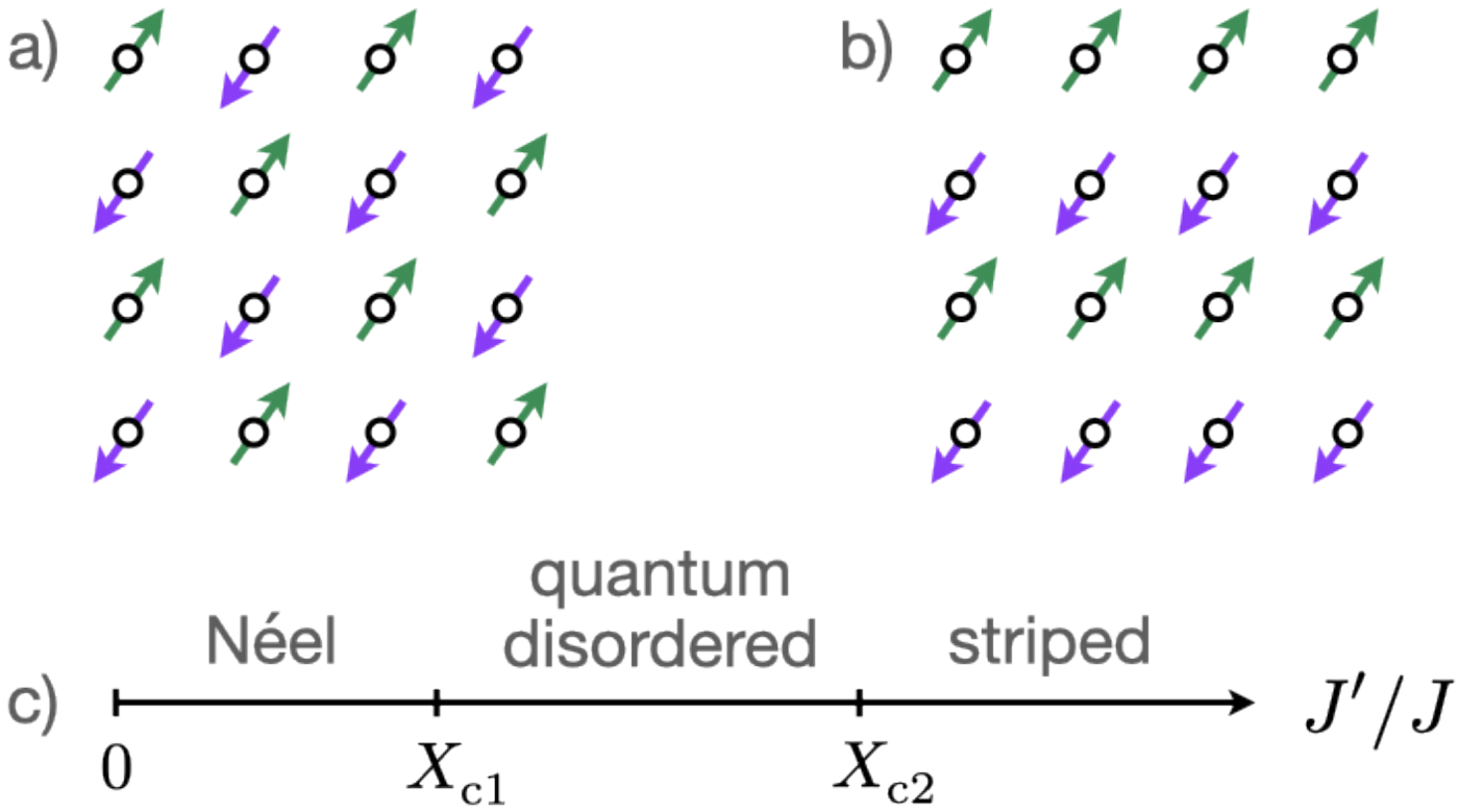}
		\caption{Forms of antiferromagnetic order  in the undoped square lattice:  a) N{\'e}el, b) Stripe AF, c) schematic phase diagram of the 2d model at large $U$ as a function of $J'/J \equiv (t'/t)^2$ }
		\label{squarex0}
	\end{center}
\end{figure}

However~\cite{balentsandjiang}, for $J^\prime/J < X_{c1} \approx 0.41$, the resulting state exhibits the same local spin correlations as the N{\'e}el state shown in Fig.~\ref{squarex0}a, and has a spin correlation length that grows exponentially with $W$, \ie\ $\xi_s \sim \exp(\alpha W)$ with $\alpha=O(1)$.  Indeed, as shown on theoretical grounds in ref.~\cite{ChakravartyPRL}, in a range of $J^\prime/J$ in which the 2d system has long-range antiferromagnetic order characterized by a renormalized spin-stiffnes $\rho_s$ and a spin-wave velocity $c$, $\alpha(J^\prime/J) = 2\pi \rho_s /\hbar c$.   Thus, the DMRG results provide compelling evidence that the 2d system has N{\'e}el order in this regime, as shown in Fig. \ref{squarex0}c.  

For $J^\prime/J > X_{c2} \approx 0.62$, the local order resembles the stripe magnetic state, shown in Fig. \ref{squarex0}b.  While we have not found data that show the systematic scaling of $\xi_s$ with $W$, $\Delta_{\rm s}$ is a sufficiently rapidly decreasing function of $W$ up to at least $W=10$ that it is reasonable to conclude that stripe LRO arises in the $W\to \infty$ limit.  On the other hand, for $X_{\rm c1} < J^\prime/J < X_{\rm c2}$, there is a spin-gap that does not  decrease substantially with increasing $W$.  These results have led to the conjecture~\cite{figandsondhi,balentsandjiang} that for this intermediate range, there is a quantum disordered phase or phases (\ie\ phases with no magnetic order) in the $W\to \infty$ limit, as shown in Fig. \ref{squarex0}c.  However, there is still some debate~\cite{shenganti} about the nature of this region in the 2d limit, \eg, whether there is valence bond crystalline order or a $\MZ\nd_2$ quantum spin liquid.

{\bf Lightly doped ladders}:
DMRG studies of the doped ladder have also primarily involved $U$ large enough that the undoped ladder is insulating, and have  focused on the range of parameters
(\ie\ $J^\prime/J < X_{\rm c1}$) which correspond to a doped N{\'e}el antiferromagnet.

Naturally, the most reliable results have been obtained for the two-leg ladder, $W=2$.  So long as $x < x_c \sim 0.3$, for the typical range of intermediate $U$ and $t'/t$, there is general consensus~\cite{noak2leg,troyer2leg,poilblanc2leg,yuvalandme} that, in the $L\to \infty$ limit, such ladders generically form a Luther-Emery liquid~\cite{lutheremery}.  Specifically, there is a non-zero spin-gap, $\Delta_{\rm s} \sim t^2/U$, a single gapless charge mode ($c=1$) and power-law equal-time CDW and SC correlations which fall as $\cos(Q\nd_{\rm CDW}r + \phi_0)|r|^{-K\nd_{\rm CDW}}$ and $|r|^{-K\nd_{\rm SC}}$, respectively.  The CDW ordering vector is equal to the value of $2k\nd_{\rm F}=W\pi n$ mandated by the generalized LSM theorem (see \S~\ref{theorems}), $Q\nd_{\rm CDW}=W\pi n\equiv 2\pi x$ and consistent with expectations from conformal field theory, $K\nd_{\rm SC} = 1/K\nd_{\rm CDW}$.  

While DMRG does not directly yield information concerning the susceptibility, once the identification with the corresponding quantum field theory is established, it follows that these same exponents can be identified with the low $T$ behavior of the corresponding susceptibilities, \ie\ $\chi\nd_{\rm CDW}(q=Q\nd_{\rm CDW}) \sim T^{-(2-K\nd_{\rm CDW})}$ when $K\nd_{\rm CDW} < 2$ and $\chi\nd_{\rm SC}(T) \sim T^{-(2-K\nd_{\rm SC})}$ if $K\nd_{\rm SC} < 2$. However, while both exponents are typically in the range $2 > K\nd_{\rm SC} > 1/2$ in which both susceptibilities diverge, the precise values of these exponents, and thus whether SC or CDW correlations are dominant, varies as a function of  $x$ and  $t'/t$.  In fact, for the two-leg ladder, the same Luther-Emery phase has been shown to exist down to $U$ as small as $U=4t$.  To get an idea of why smaller values of $U$ are so difficult, note that for $t'=0$ and $x=1/12$, the spin correlation length $\xi_s \approx 30$ lattice constants;  it is thus found that  ladders up to $L=200$ must be treated to get reliable results~\cite{yuvalandme}.  (Interestingly, in this case, 
the CDW and SC susceptibilities are equally divergent, $K\nd_{\rm SC}\approx K\nd_{\rm CDW} \approx 1$.)

\begin{figure}[t]
	\begin{center}
		\includegraphics[scale=0.4]{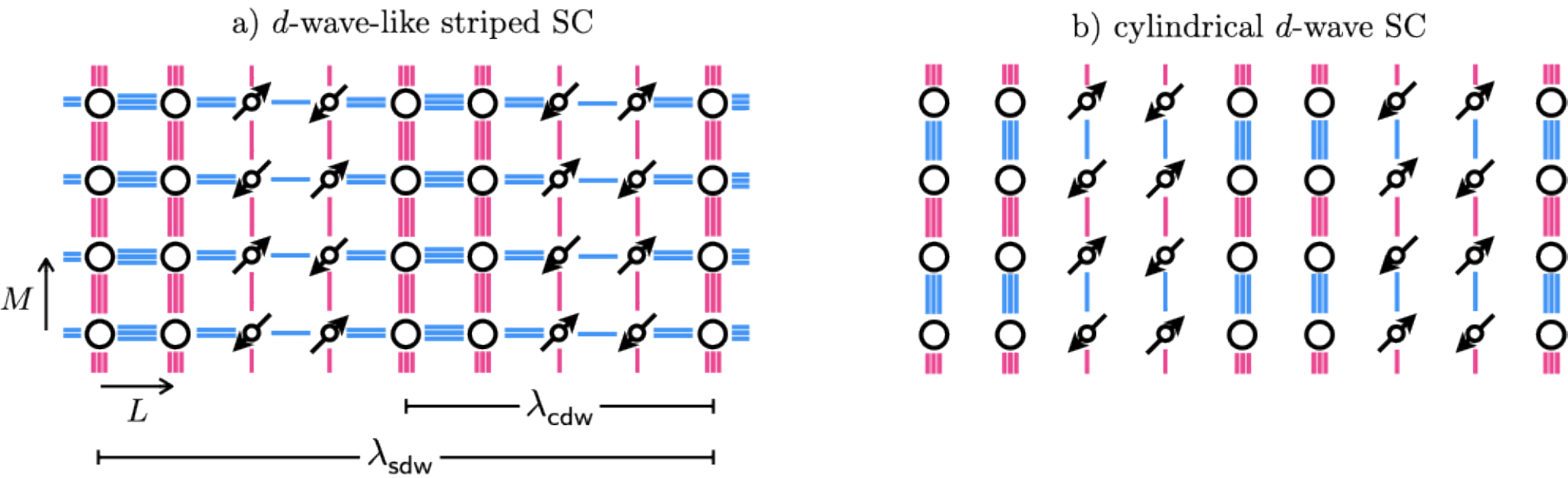}
		\caption{Various patterns of intertwined short-range order for the square lattice Hubbard model at moderate $x$ and intermediate $U$.  Here the size of the circles indicates represents the doped hole density, the arrows represent the site magnetization, and the thickness of the lines the magnitude of the singlet pair field on each bond;  the sign of the pair-field is color coded with red positive and blue negative.  Here   a) Represents coexisting SDW, CDW, and $d$-wave-like SC, b) (relevant in particular to the $M=4$ leg cylinder) represents SDW, CDW, and cylindrical ``true $d$-wave'' SC, and c) represents SDW, CDW, and PDW order. }
		\label{stripes}
	\end{center}
\end{figure}

The nature of the short-range correlations on the two-leg ladder are also interesting.  Although the SDW correlations decay exponentially with distance, the spin correlation length is sufficiently long that one can identify a preferred ordering vector, $Q\nd_{\rm SDW}= \pi n \equiv \pi \pm \pi x$.  Importantly, this is mutually commensurate with the CDW ordering vector, $Q\nd_{\rm CDW} \equiv 2Q\nd_{\rm SDW}$. Also, the superconducting order is ``$d$-wave-like'' in the sense that the pair-field correlations on bonds parallel to and perpendicular to the ladder direction have opposite signs.  (Since the ladder does not have any symmetry relating parallel and vertical bonds, this is not a precise symmetry classification, and indeed the magnitude of the pair-field correlations are somewhat different on the two sets of bonds.)  The  short-range order is thus a two-leg version of the correlations shown schematically in Fig. \ref{stripes}a, where for graphical simplicity we have drawn this picture for the commensurate case $x=1/4$ so that $\lambda\nd_{\rm CDW} =4$.  The relative phase of the CDW, SDW, and SC modulations are representative of what is seen in the calculations, \ie\ the spin order is strongest where the SC order is weakest and where the local doped hole concentration is closest to 0.  (The fact that the CDW is ``bond-centered,'' \ie\ that it preserves reflection symmetry about a bond-centered mirror plane, was likewise chosen for
purposes of illustration only.)

Even for the two-leg ladder, there are circumstances in which other phases arise.  As already mentioned in \S \ref{strongcoupling}, for large enough $U$, for $0< x \lesssim 0.2$, there is a fully polarized (half-metallic) phase - a result that was also established using DMRG~\cite{liuferro}.  Moreover, for special rational values of $x$ (such as $x=1/8$) and for a range of $U/t$, commensurate CDW long-range order is possible~\cite{affleckwhite,schulzcommensurate}, \ie\ a discrete breaking of the translational symmetry.  The ordering vector in this case is, as before, $\lambda\nd_{\rm CDW} = 2/x$, but now there is a gap to charge excitations as well a spin-gap.

Moving to $W=4$ leg ladders and cylinders, a still richer variety of behaviors has been observed~\cite{whitescalapino4leg,whitescalapinotprime,edwinnpj}.  In particular the nature of the states for a given value of $x$ is found to depend significantly on the value of $t'$. 
Under some circumstances, especially in 4-leg cylinders with $t'$ small and negative,\footnote{In DMRG studies, a value of $t' \approx - 0.3t$  is often taken as a value representative of the band-structure of the hole-doped cuprates.}
there appears to be~\cite{jiangandtom,jiangandme,whitenew} a Luther-Emery liquid phase with $c=1$, divergent CDW and SC susceptibilities with the product $ K\nd_{\rm CDW} K\nd_{\rm SC}=1$ and with the $K_{\rm SC}$ in the range $0.5 < K_{\rm SC} \lesssim 2$, and exponentially falling SDW correlations with $\xi_{sdw} \sim 10$. Where this occurs, it is typically also true that $Q\nd_{\rm CDW} = 4\pi x$.  The sketch in Fig. \ref{stripes}b is a caricature of the short-range order seen here.  Recalling that periodic boundary conditions have been enforced around the cylinder, it is apparent that the SC order -- which oscillates in sign between neighboring bonds around the cylinder -- has a ``true $d$-wave'' form~\cite{dodarojiang,whitenew} in the sense that it changes sign under a $C_4$ rotation about the central axis of the cylinder.  Such a state is likely to be a peculiarity  of the 4-leg cylinder.

For   $t'= 0$, the $W=4$ leg cylinder shows significantly different tendencies.  The most extensive studies~\cite{dodarojiang,whitesimons,jiangandtom,whitenew}  have been carried out primarily for the (assumed representative) value of $x=1/8$.  Here the preferred CDW ordering vector is half what it was in the previous case, $Q\nd_{\rm CDW} = 2k\nd_{\rm F} \equiv 2\pi x$.  Moreover, it seems that the SC correlations fall rapidly -- most probably exponentially - with distance.  (However, in this case, the SC correlation length, $\xi\nd_{\rm SC}$ is remarkably long - it is estimated in ref.~\cite{jiangzaanen} that $\xi\nd_{\rm SC} \approx 18$). There is again significant short-range SC and SDW order, with, as in Fig.~\ref{stripes}a, the SC order being $d$-wave-like, and the SDW ordering vector satisfying the mutual comensurability condition, $2Q\nd_{\rm SDW} \equiv Q\nd_{\rm CDW}$.  This, if extrapolated to the 2d limit, is suggestive~\cite{whitesimons} of a commensurate unidirectional CDW insulating state, either with or without accompanying SDW order, with a significant but strictly short-range correlated tendency toward $d$-wave SC.  Thinking of the peaks in the CDW as being a stripe of high doped hole density, this state, which arises naturally in Hartree-Fock calculations in 2d, is sometimes referred to as having ``full stripes'' -- full, in the sense that since $\lambda\nd_{\rm CDW} = 1/x$, there is one doped hole per site along the length of each stripe.  By contrast, the state with $\lambda\nd_{\rm CDW} = \frac{1}{2x}$ ($Q\nd_{\rm CDW} = 4\pi x$) is then referred to as ''half-filled'' stripes.

Patterns of local order corresponding to partially-filled stripes of different varieties have been found in calculations on $W=6$ 
leg ladders~\cite{2jeckelmanstripes,whiteandscalapino6}. 
Very recently, there have been the first DMRG studies\cite{jiang2021high,gong2021robust,jiang2021ground} (actually, of the $t-J$ model) to find direct evidence of (quasi)-long-range  d-wave-like SC order in cylinders with $W>4$. Not surprisingly, the SC correlations tend to be strongest for values of the parameters (e.g. $t^\prime/t$) for which the CDW correlations are relatively weaker\cite{gong2021robust,jiang2021ground}.  Generally, the phases with strong SC correlations occur only for $x$ greater than a (parameter dependent) critical value of $x$ - the one exception to this\cite{jiang2021high} being the case of very large $t^\prime \approx t/\sqrt{2}$, where the state at $x=0$ is (as already discussed) a quantum para-magnet (probably a spin liquid), and where SC quasi-long-range-order appears down to the smallest values of $x$ ($x=1/18$) explored so far.  
A periodic modulation of the hopping was found to significantly enhance the d-wave like pairing correlations, as well~\cite{Jiang2021stripe}. 
As these are very new results,  it is still necessary for further studies to be carried out to determine the full phase diagram.

\subsubsection{The honeycomb lattice} DMRG studies have been performed for the Heisenberg model on the honeycomb lattice for a range of $J'/J$~\cite{Gong2013,Gong2015}. The behavior of the honeycomb model appears to have similar motifs as the more studied square and triangular lattices. In particular, there is evidence for an intermediate spin-disordered phase for a range of $J'/J$ between the N{\'e}el and striped phases. All or much of this intermediate phase seems to possess plaquette order~\cite{Gong2015}. Some DMRG studies have also been performed on the honeycomb Hubbard model away from half filling~\cite{Jiang2014,Yang2021,Qin2021}. We will not further summarize these results here.

\subsubsection{The triangular lattice} DMRG studies of  the Hubbard model on the triangular lattice have recently produced evidence of a number of intriguing behaviors~\cite{hongchentriangle,zaleteltriangle,ashvintriangle,shengchiraltriangle,shengtriangle,eggerttriangle}.  Unfortunately, there is still considerable disagreement in the inferences drawn from different studies, making it difficult to give a definitive review.

{\bf Undoped cylinders}   with $3 \leq W \leq 5$ have been studied at $x=0$. There is strong evidence (and a general consensus) that there are at least three distinct phases as a function of $U/t$.  For $U< U_{\rm c1} \approx 8t$ there is a compressible phase that has been identified as ``metallic'' in several DMRG studies. More specifically the DMRG studies have concluded that it is metallic in the sense of being adiabaticaly connected to the non-interacting limit, \ie\ that there are gapless spin and charge modes corresponding to each Fermi surface crossing of the non-interacting band-structure.  This conclusion is difficult to reconcile with  a weak coupling analysis~\cite{yuvalandme} of the sort described in \S \ref{weak}.  Rather, assuming there are no as yet undetected transitions with a critical value of $U<U_{\rm c1}$, it is likely that most of these modes are gapped, but that the gaps are sufficiently small (the correlation lengths sufficiently large) that they have been overlooked in DMRG studies to date.  Indeed, the weak coupling analysis suggests that the small $U$ phase is a Luther-Emery liquid ($c=1$) with time reversal symmetry breaking - \ie\ an extension of the $d+id$ SC phase that arises in 2d at small $U$ (see \S \ref{weak}) to a cylinderical geometry.

The most intriguing phase at $x=0$ occurs for $U_{\rm c1} < U < U_{\rm c2} \approx 10t$.  Here there is evidence that the system becomes a spin liquid in the 2d limit.  For instance, for $M=4$ and suitable boundary conditions, there seems to be a fully gapped state with only very short-range spin correlations.  It was concluded in ref.~\cite{zaleteltriangle} that this phase is chiral - indicating that in the 2d limit it would correspond to a time reversal symmetry breaking chiral spin liquid
(\ie\ a Kalmeyer-Laughlin phase~\cite{kalmeyerlaughlin}).  This has a degree of naturalness as it is easy to conceive
a continuous transition from a $d+id$ SC to a chiral spin liquid.  However, it should be mentioned that other studies~\cite{shengtriangle,shengchiraltriangle} have found evidence that this intermediate phase is a fully gapped but non-chiral state -- which 
is suggestive that there is a $\MZ\nd_2$ spin liquid (\ie\  a Moessner-Sondhi phase~\cite{moessnersondhi}) in the 2D limit -- or even a gapless spin liquid~\cite{eggerttriangle} with a nodal spinon spectrum.  

Finally, for $U_{\rm c2} < U$ there is a phase with local correlations corresponding to the 120$^\circ$ antiferromagentic state, known to be the ground state of the 2d model in the large $U$ limit.

{\bf Doped cylinders}: The nature of the lightly doped system clearly depends on the character of the undoped parent state.  The resulting phase diagram  has not been nearly as thoroughly investigated as in the case of the square lattice.    Particularly interesting is the  fate of a lightly doped spin liquid. A spin liquid can be thought of as a quantum disordered superconductor, such that upon light doping, the ``pairing'' is a property inherited from the insulating state while the superfluid stiffness (and hence the ordering temperature) vanishes continuously as $x \to 0$~\cite{anderson87,rokhsarandme,laughlinsc}. Specifically, light doping of a gapped %$\MZ\nd_2$ 
 spin liquid could lead  to a  gapped superconductor - with the ``superconducting'' gap derived from the spin-liquid gap -  which is chiral ($d+id$)~\cite{rokhsardid} or non-chiral depending on the nature of the spin-liquid from which it arose - while a nodal spin liquid naturally connects to a nodal superconductor~\cite{balentsandnayak}.  
There is preliminary DMRG evidence that each of these may occur on  the triangular lattice under appropriate circumstances~\cite{jiangdopedsl,ashvintriangle,eggerttriangle,hongchentriangle,eunahdopedtriangle}.

\subsubsection{The kagome lattice} The Heisenberg model on the kagome lattice is a paradigmatic example of a  
geometrically frustrated quantum magnet. It has long been suspected that this system does not order magnetically, even with only nearest-neighbor $J$.  Different ground states have been proposed based on exact diagonalization of finite clusters~\cite{Zeng1990,Chalker1992,Lecheminant1997,Lauchli2011,nakano2011numerical,Lauchli2019}, 
and various approximate approaches. [For a discussion, see Refs.~\cite{Lauchli2019,fradkinkagome}.] 

DMRG calculations on cylinders of width up to $W=17$ have been performed~\cite{Yan1173,Jiang2008,jiang2012identifying,Depenbrock2012,He2017}. (In some of these calculations, a small second-neighbor $J'\le 0.15J$ has been applied.) These works consistently find a quantum disordered ground state with no sign of magnetic or valence bond crystalline order. This suggests a quantum spin liquid ground state. Unfortunately, there is disagreement on the nature of this spin liquid, \eg, whether it has a spin gap~\cite{jiang2012identifying} or a gapless Dirac spinon spectrum~\cite{He2017}. The disagreements suggest that the long-range correlations may not yet be fully converged with respect to the bond dimension, and more extensive studies are needed to resolve the nature of the ground state. 

In the  studies of the doped system carried out to date~\cite{Jiang2017,hongchenkagome}, light doping of the kagome spin liquid seemingly leads to  a ``holon'' crystal phase - an insulating state with one doped hole but zero spin per unit cell - rather than to a superconductor.

\subsection{Quantum Monte Carlo results} The determinant quantum Monte Carlo method~\cite{Blankenbecler1981,Hirsch1983,White1989,assaad2002quantum} is a powerful technique to find equilibrium properties of interacting fermions. It is controlled, in the sense that the results are guaranteed to converge to the exact answers upon increasing the computational effort. Unfortunately, in many cases of interest, the applicability of the DQMC method is limited due to the fermion sign problem~\cite{White1989a}. In these cases, the computational cost for a given accuracy increases exponentially with the system size and the inverse temperature. The repulsive Hubbard model suffers from the fermion sign problem in DQMC at generic values of the filling. Below, we briefly review some useful results that were nevertheless obtained using DQMC, either for special parameters where the sign problem is absent, or by employing very large computational resources to overcome the sign problem. 

The $U>0$ Hubbard model is sign problem free if the system is particle-hole symmetric~\cite{White1989}. This is the case at half filling ($n=1$) on a bipartite lattice (\eg, square or honeycomb), when the intra-sublattice hopping is set to zero. Note that the sign of $U$ can be changed by performing a particle-hole transformation on one spin species, showing that the $U<0$ Hubbard model is sign problem free with zero total magnetization and an arbitrary filling. The Hubbard model has been simulated using DQMC on a square lattice~\cite{White1989}, showing a clear Mott gap and an antiferromagnetic (N\'eel) ground state for all $U>0$. 

On the honeycomb lattice, the semimetal state with Dirac nodes is stable for sufficiently small $U$, whereas the ground state for sufficiently large $U$ is a collinear sublattice antiferromagnet with opposite spin polarization on the two sublattices~\cite{sorella1992semi,Paiva2005}. Initial simulations suggested a quantum spin liquid state at intermediate $U$ between the semimetal and the AF~\cite{meng2010quantum}; however, the current consensus is that there is a direct continuous transition between the semimetal and AF phases with no intermediate phase~\cite{sorella2012absence,Assaad2013}.

The bilayer Hubbard model has been studied at half filling for the square~\cite{Bulut1992,Scalletar1994,Bouadim2008} and honeycomb~\cite{Lang2012,pujari2016interaction} lattices. In the latter case, motivated by Bernal-stacked bilayer graphene, an AB stacking has been considered. In this case, the non-interacting band structure has quadratic band touchings, and the system may be expected to be unstable even in the presence of arbitrarily weak $U$, as discussed in \S \ref{weak}. However, it was found that for weak interactions, the dispersion gets renormalized, and each quadratic band touchings splits into four linearly dispersing Dirac points~\cite{pujari2016interaction}. As a result, a finite $U$ is needed to bring the system from the semimetal into the AF phase, much as in  the single layer honeycomb model. 

Finally, DQMC has been applied to the square~\cite{huang2017numerical,huang2019strange,edwinnpj,liu2021tendencies} and honeycomb~\cite{zhu2019quantum} lattice away from half filling. These calculations require massive computing resources. The square lattice calculations have currently been performed for systems for size $L=8$ at temperatures down to $T/t \simeq 0.2$ and interaction strength $U/t = 8$. At these temperatures, finite size effects are found to be small, such that $L=8$ is probably sufficient 
to make inferences about the thermodynamic limit. No symmetry broken phase was found. However, there are significant short-range correlations~\cite{huang2017numerical} indicative of incommensurate unidirectional spin density wave,
$d$-wave superconductivity, as well as nematic bond order~\cite{liu2021tendencies}, that grow as the temperature decreases. 

\section{Are there exotic phases in the Hubbard model?}

Progress has been made in recent years in constructing ``reverse engineered'' model Hamiltonians - such as the Kitaev model or the quantum dimer model - that exhibit exotic quantum phases of various sorts.  This is sufficient to settle certain long-standing issues of principle - such as whether quantum-spin-liquid phases exist.  However, to address whether it is reasonable to expect such phases to arise without extreme fine tuning, it is worthwhile to ask whether these arise in some version of the Hubbard model - with some particular lattice geometry or some range of $U/t$ or $t^\prime/t$.  
\subsection{Phases with exotic broken symmetries}
Two novel classes of broken symmetry phases have been proposed as candidate orders to account for some observed anomalies in the cuprate high temperature superconductors:  states with orbital loop current order~\cite{Varma1999,chakravartylaughlin} and pair-density-wave states~\cite{himeda,layerdecoupling,pdwreview}  with spatially modulated superconducting order.

\subsubsection{Orbital Loop current order}  These are states that are close relatives of a charge density wave, but which break time-reversal symmetry, \ie\ in which
\begin{align}
   i\sum_{\sigma} \langle c\yd_{j,\sigma} c\nd_{k,\sigma} - c\yd_{k,\sigma} c\nd_{j,\sigma}\rangle = J\nd_{jk} \neq 0\quad.
\end{align}
Two particularly important such proposals are a zero-momentum ``intra-unit cell orbital antiferromagnet'' in which $J_{jk} = J(\BR_j-\BR_k)$, which has ordering vector $\BQ = \boldsymbol{0}$, and a ``$d$-density-wave,'' in which - on a square lattice in particular - $J_{jk} = e^{i\BQ\cdot \BR_j} J(\BR_j-\BR_k)$ with $\BQ=(\pi,\pi)$.  In both these examples, $J(\BR)$ transforms non-trivially under the operations of the point group. Another related state - of which  the ``triplet d-density wave'' is an example~\cite{tripletddw} -  is a spin-current density wave -  which preserves time-reversal symmetry.

As far as we know, convincing evidence for the existence of such  phases has not been found in either numerical or controlled approximate studies of Hubbard models.  DMRG calculations on doped Hubbard ladders that have probed such order typically have found very weak, extremely short-range current-current correlations~\cite{whitescalapinoloops,jeckelmannloops,devereauxloops,dodarojiang}\footnote{In \cite{jeckelmannloops}, indications of loop current order were found for relatively large $x$ in a three band model with substantial nearest-neighbor $V$, but this was found to arise in an ``unphysical'' range of parameters, i.e. one in which the undoped system is not an antiferromagnetic insulator.}. 

There is evidence (discussed in \S \ref{topological}) that a chiral spin-liquid phase arises in a small but finite range of $U/t$ in the undoped Hubbard model on the triangular lattice.  Moreover, for a band-structure with a symmetry protected quadratic band-touching, there is a dominant tendency toward an anomalous quantum Hall state at weak coupling~\cite{quadratic}. Although these states break time reversal symmetry, they must have a zero orbital current on any link lying in a mirror plane. However, a CDW state in doped versions of either of these seems likely to support orbital loop current order.

\subsubsection{Pair Density wave}
A pair-density-wave (PDW) is a close relative of the famous FFLO states \cite{FF1964,LO1964} that arise (at least in theory) in conventional superconductors in response to a small degree of spin-polarization.  As the name suggests, this is a superconducting state with a pair-field that is spatially modulated:
\begin{align}
    \big\langle\, c\yd_{j,\uparrow} c\nd_{k,\downarrow} + c\yd_{k,\uparrow} c\nd_{j,\downarrow}\,\big\rangle = \Phi(\BR\nd_j,\BR\nd_k) \neq 0\quad,
\end{align}
where in contrast with any conventional superconducting state, the spatial average of $\Phi$ vanishes, $\sum_{\BR} \Phi(\BR, \BR+\Br) =0$ for fixed $ \Br$.
(This is a spin-singlet PDW - one can also consider the possibility of a spin-triplet PDW.)  Such states have been found as close competitors in certain variational treatments of the Hubbard model, and short-range PDW correlations have been reported in several different DMRG calculations on Hubbard ladders.  Moreover, to date, no clear evidence of PDW long-range order (or even of sufficiently strong quasi-long-range order to lead to a divergent PDW susceptibility) has been presented in studies of any version of the Hubbard model~\cite{White2009,tomstalk}, except in one dimension~\cite{Berg2010,Jaefari2012}.  However, the observation of clear PDW short-range order in some cases is a promising point of departure for future investigations~\cite{himeda,corbozrice,leepdw}.

\subsection{Phases with topological character}
\label{topological}
There has been an extensive search for spin-liquid phases in Hubbard models with an odd integer number of electrons per unit cell.  Here, as discussed in \S~\ref{theorems}, it follows from the generalized LSMOH theorem that any insulating phase that  preserves translation symmetry must be a topologically ordered ``spin liquid'' phase that cannot be adiabatically connected to any band-insulator. At present, while there is no absolutely convincing evidence of such a phase, as discussed in \S \ref{DMRG}, there are several systems for which DMRG results are highly suggestive of the existence of quantum spin liquid phases, albeit for relatively narrow ranges of parameters.

\section{Speculations concerning the phase diagram in $d \geq 2$}

Until this point, we have presented results that are established with some degree of theoretical certainty. Alas, this means we have not been able to present any results for what might a-priori be considered the most important regime - spin-$\half$ fermions with density near but not equal to 1 per site and $U$ comparable to the band-width. This is the range of parameters that 
is likely most relevant to phenomena in a host of highly correlated materials including the cuprate and various organic superconductors.  In this final section we make a speculative attempt to extrapolate from regimes in which our theoretical 
understanding is solid to obtain a global picture of the phases and regimes of the Hubbard model - particularly focused on the properties of the  model in 2d.

\subsection{The square lattice Hubbard model}

\begin{figure}[t]
	\begin{center}
		\includegraphics[scale=0.35]{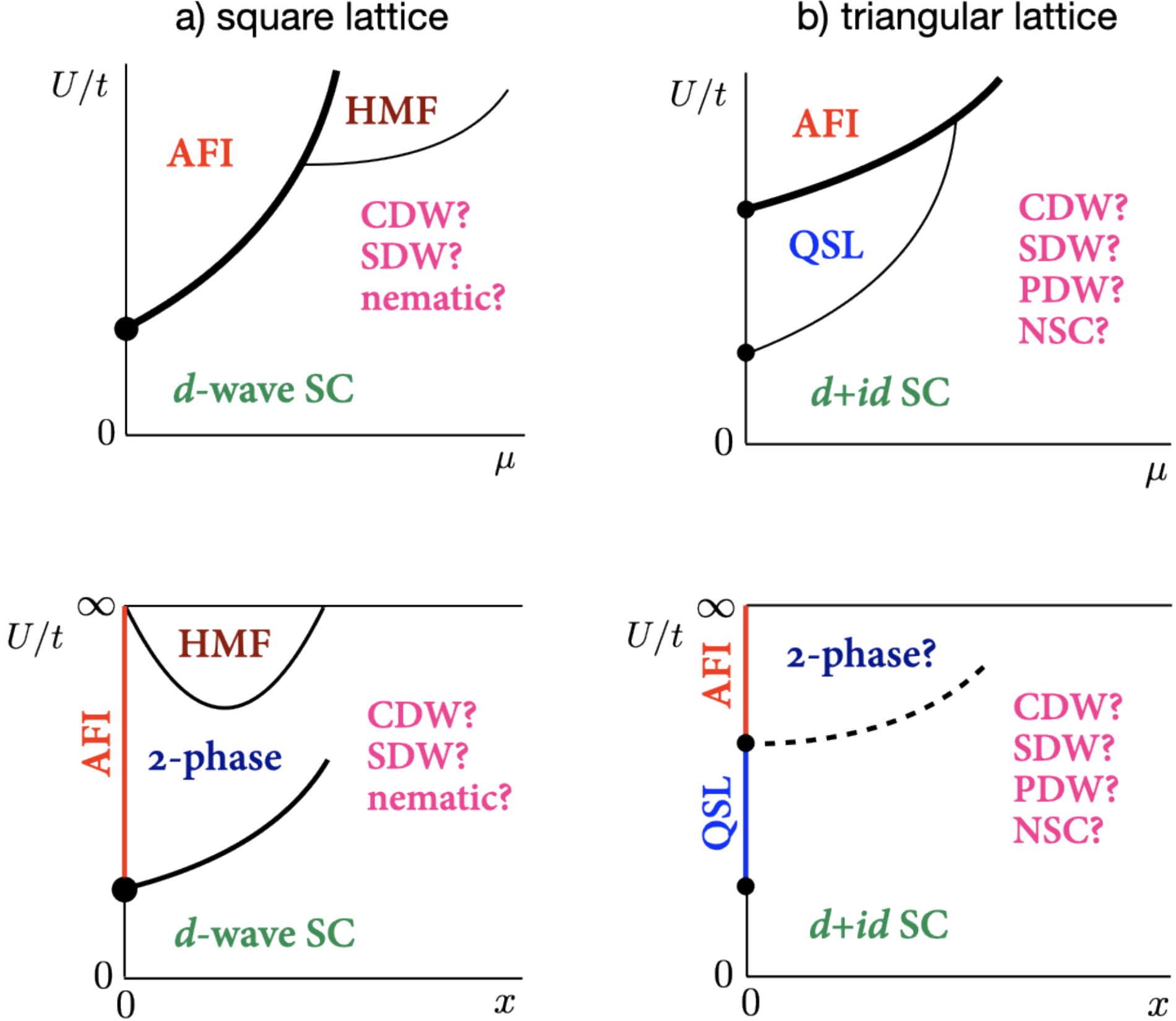}
		\caption{Speculative ground state phase diagrams of the Hubbard model as a function of $U$ and chemical potential, $\mu$ (upper panels) or doped hole density, $x$ (lower panels). {\bf a) Square lattice with  $t \gg |t'| >0$:}
	AFI indicates an insulating (incompressible) phase with  N\'eel AF order and doped hole-density $x=0$. The thick solid line indicates a first-order transition from the AFI to a compressible phase,  and the black circles denote points at which $x=0$ in the compressible phase. ``2-phase'' denotes a region of two-phase coexistence.  HMF denotes a half metallic ferromagnetic phase. 
		 At intermediate $U/8t$, there is clear numerical evidence of multiple local ordering tendencies of comparable strengths including unidirectional CDW, colinear unidirectional SDW, nematic, and $d$-wave SC. Which of these phases actually orders is still uncertain. 
		 {\bf b) Triangular lattice:} 
		 At weak coupling, the ground state is a $d_{x^2-y^2}+id_{xy\nd}$ SC. DMRG studies suggest that for $x=0$ as a function of increasing $U$ there is an insulating quantum spin liquid (QSL) phase when $U$ is comparable to the bandwdith, $9t$, that lies between the small $U$ SC and the large $U$ three-sublattice $120^\circ$ ordered AFI phase. The transition from the QSL to the AFI is likely first order. For $U\sim 9t$ and finite $x$, there is DMRG evidence  of at least short-range tendencies to $d+id$ and nematic (NSC)superconducting order, as well as CDW, SDW and PDW phases. 
	}  
		\label{squarePD}
	\end{center}
\end{figure}

In Fig. \ref{squarePD}a we show a schematic  $T=0$ phase diagram in the $U/t - \mu$ plane of the  square lattice Hubbard model with non-vanishing second neighbor hopping matrix, $t^\prime/t \neq 0$.  On the basis of the weak coupling analysis, we know that for $U/8t\ll 1$,  there is a uniform $d$-wave SC phase with no other forms of coexisting order. (Here $8t$ is the non-interacting bandwidth.) In the large $U$ limit and for a range of doping $0<x\lesssim 0.2$, one can conclude on the basis of DMRG studies that  there is a fully polarized ferromagnetic phase - a ``half-metallic ferromagnet'' (HMF).  Moreover, there are compelling theoretical arguments~\cite{ekl,altshulerandhuse} (which we will not review here) that there is a direct first-order transtion from the AFI to the HMF phase, giving rise to the region of two-phase coexistence indicated, with the density of the compressible phase going as $x_c \sim \sqrt{8t/U}$ as $U\to \infty$.

Independent of $U$, in the dilute electron limit (not shown) the system behaves as a Fermi liquid behavior down to an exponentially low energy scale, below which it presumably forms an unconventional superconducting state by some version of the Kohn-Luttinger mechanism.  Finally, for the half filled band ($x=0$) and  $U/8t$ greater than a critical value,  $\alpha_c $,  there is an insulating phase which, so long as $t^\prime/t\lesssim 1/2$  (so the magnetism is not terribly frustrated), exhibits long-range N\'eel order. (As discussed in \S \ref{hartreefock}, $\alpha_c \sim 1/\ln|t/t^\prime|\to 0$ as $t^\prime/t \to 0$.)

In the figure, we have indicated by the black circle a presumed direct (and consequently first order) transition from the $d$-wave superconductor at $x=0$ to the insulating antiferromagnetic phase.  Such a direct transition occurs in Hartree-Fock approximation (or equivalently in a suitable large $N$ limit) for small $t'/t$, although at larger $t'/t$ there can arise an intermediate antiferromagnetic metal (\ie\ only partially gapped) phase~\cite{unpublished}. Assuming the transition at $x=0$ is first order, the transition must remain first order for a range of chemical potentials.  Consequently, for small $U/8t> \alpha_c$  there must exist a two-phase coexistence region corresponding to the thick solid line in the figure.

Indeed, arguments suggesting that the antiferromagnetic insulating phase is generically bounded by a line of first order transitions (leading to phase separation) were summarized in ref.~\cite{SKLosAlamos}, and more recently have been supported by extensive variational Monte Carlo studies of the Hubbard model in ref.~\cite{sorella}.  Another candidate small $x$ phase when $U/8t\gtrsim \alpha_c$ that is suggested by Hartree-Fock studies~\cite{zaanenstripes,shultzstripes1,machidastripes1} (especially in the limit $t'=0$) is an insulating  incommensurate unidirectional colinear SDW (stripe) phase.  The stripe phase can be thought of as a form of micro-phase-separation.

Unfortunately, the structure of the middle part of the phase diagram, where $U/8t \sim 1$ and $ 1/12 \lesssim x \lesssim 1/3$, is presently unsettled.  From DMRG and other studies, it is clear that there are strong local tendencies toward $d$-wave SC, as well as unidirectional (stripe) CDW and SDW orders.  Implicit in the observation of striped states is a strong tendency toward lattice rotational symmetry breaking, \ie\ nematic order\footnote{This interpretation requires some care, because of the effects of the boundaries, present in most DMRG calculations.}.  

That all DMRG studies (as well as other less controlled methods)  find   strong local tendencies toward $d$-wave SC  is compatible with the supposition that the Hubbard model at intermediate coupling captures an essential feature of cuprate physics that leads to high-temperature $d$-wave SC.  However, whether the Hubbard model has more than  a ``$d$-wave tendency'' - \ie\ whether it actually supports a robust $d$-wave SC phase at intermediate $U$ or not - is still unsettled.  Under most circumstances, especially in broader ladders,  the DMRG studies suggest that the competing tendency toward CDW order may be stronger.

\subsection{The triangular lattice Hubbard model}
Fig.~\ref{squarePD}b  shows a conjectured phase diagram for the triangular lattice Hubbard model.  Again, from a weak coupling analysis, we know that the small $U$ portion of the phase diagram is superconducting, likely a $d+id$ SC~\cite{nandkishore2012chiral,Nandkishore2014}.  At $n=1$ and large enough $U$, the model is equivalent to a spin 1/2 Heisenberg model, for which strong evidence exists that the ground state has long-range, coplanar three sublattice insulating antiferromagnetic order - the ``120$^\circ$ state.''  DMRG studies consistently indicate the existence of an intermediate insulating phase without any long or quasi-long-range antiferromagnetic order - corresponding to some sort of quantum spin liquid (QSL).  Thus, at $x=0$ there are two transitions (indicated by the solid circles) as a function of $U/8t$.  The exact nature of the QSL, however, is still contravertial.

There is some evidence that the QSL in question is a chiral spin liquid, which would  be compatible with a continuous transition to a chiral ($d+id$) SC upon light doping, as shown in the figure.  Then at larger doping and intermediate values of $U/8t$, there is evidence from DMRG studies of a local tendency toward a variety of possible broken symmetry phases including SDW, CDW, PDW and nematic SC.  However, again, which of these orders exist as ground state phases in 2d is still an open question. Preliminary evidence~\cite{hongchentriangle} that the SC tendencies are particularly strong at small $x$ and for $U$ in the spin-liquid regime offers some encouragement for the long cherished ideal that a spin liquid may be a high temperature superconductor waiting to happen.

\section{Important Open Questions}
We end by highlighting some of the major outstanding challenges in the physics of the Hubbard model. 

{\bf Is the Hubbard model a high temperature SC?}  It has been established that the repulsive $U$ Hubbard model has a superconducting ground state at small $U$, but it still remains uncertain whether - and if so under what circumstances - it supports ``high temperature superconductivity.''  In other words, are there circumstances (\ie\ some range of band-structure parameters) in which the Hubbard model is superconducting when all the energy scales are comparable, \ie\ when $U$ is on the order of the bandwidth, so that $T\nd_{\rm c}$ - if a SC state arises - is a sizable fraction of $E\nd_{\rm F}$. To get a quantitative feeling, recall that the bandwidth in the cuprates is of the order of $W \sim 2$eV. Taking $W \simeq 8t$, we obtain that a $T_c$ of $0.05t$ would correspond to $T_{\rm c}\sim 150$K, of the order of the maximal transition temperature found in the cuprates.

While affirmative answers to this question have been suggested on the basis of various approximate calculations, the presence of multiple intertwined orders - with the consequent existence of subtle energies that are difficult to capture reliably - renders the validity of these approximate results uncertain.

{\bf What  exotic phases arise?} At the same time, while there are various reasons to feel that exotic forms of quantum order can arise in the Hubbard model - at least if the underlying band-structure cooperates - currently the evidence for these states ranges from suggestive to absent. Ideally, one would like to identify versions of the model that unambiguously exhibit various forms of insulating quantum spin liquids, pair-density-wave SCs (in more than one dimension), and/or possible forms of orbital loop current order.  

{\bf What sort of non-Fermi-liquid behaviors occur at elevated $T$?}  We have hardly touched on the nature of the model at finite $T$ (although much is known) and have totally neglected any issues associated with near equilibrium dynamical properties, much less the far from equilibrium behaviors that are of so much recent interest.  For instance, for $U$ on the order of the bandwidth, there most probably will not be a broad range of $T$ above all ordering temperatures in which the system can be well described by the weakly interacting quasi-particles of Fermi liquid theory.  What the behavior is  in this regime of $T$  - especially including what processes govern the dissipative linear response of the system - is one of the most important open areas in the field.

{\bf Have we learned anything useful?}  Finally, we have discussed controlled solutions of the Hubbard model - and have emphasized the limited progress that has been made even on this simplest of all model strongly correlated systems.  Obviously, from a broader perspective, what is needed are much simpler and more versatile methods of solution~\cite{gul,DMFTreview,Simons2015} that once benchmarked by comparison with the controlled results discussed here, can be applied more widely - perhaps even in ways that interface with microscopic electronic-structure approaches~\cite{kent2018toward,paul2019applications,jarrell}.

\vskip 5mm

\section*{Acknowledgements}

It is a pleasure to acknowledge numerous extremely helpful discussions about the Hubbard model and related topics with too many colleagues to list.  In particular, however, the writing of this paper was greatly assisted by input from Ian Affleck, Assa Auerbach, Vladimir Calvera, Andrey Chubukov, Youjin Deng, Tom Devereaux, Eduardo Fradkin, Zhaoyu Han, Hong-Chen Jiang, Elliott Lieb, Dror Orgad, Mohit Randeria, Doug Scalapino, and Richard Scalettar. SAK was supported, in part, by the National Science Foundation (NSF) under Grant No. DMR2000987. EB acknowledges support from the European Research Council (ERC) under grant HQMAT (Grant Agreement No. 817799). SR was supported in part by the Department of Energy, Office of Basic
Energy Sciences, Division of Materials Sciences and Engineering, under contract DE-AC02-
76SF00515.

\bibliographystyle{unsrt11}

\bibliography{Hubbard_references.bib}
\end{document}